\documentclass[%
 reprint,prb,
superscriptaddress,
nofootinbib,
 amsmath,amssymb,
 aps,
]{revtex4-2}
\usepackage{graphicx}% Include figure files
\usepackage{dcolumn}% Align table columns on decimal point
\usepackage{bm}% bold math
\usepackage{hyperref}% add hypertext capabilities
\usepackage{xcolor}
\usepackage[normalem]{ulem}
\begin{document}

%\preprint{APS/123-QED}

%\title{Plasmons and long-wavelength density fluctuations in the strange metal Bi$_2$Sr$_2$CaCu$_2$O$_{8+x}$}

\title{Consistency between reflection M-EELS and optical spectroscopy measurements of the long-wavelength density response of Bi$_2$Sr$_2$CaCu$_2$O$_{8+x}$}

\author {Jin Chen}\email{jinc3@illinois.edu}
\affiliation{Department of Physics and Materials Research Laboratory, University of Illinois at Urbana–Champaign, Urbana, IL 61801, USA}
\author {Xuefei Guo}
\affiliation{Department of Physics and Materials Research Laboratory, University of Illinois at Urbana–Champaign, Urbana, IL 61801, USA}
\author {Christian Boyd}
\affiliation{Department of Physics and Materials Research Laboratory, University of Illinois at Urbana–Champaign, Urbana, IL 61801, USA}
\author {Simon Bettler}
\affiliation{Department of Physics and Materials Research Laboratory, University of Illinois at Urbana–Champaign, Urbana, IL 61801, USA}
\author {Caitlin Kengle}
\affiliation{Department of Physics and Materials Research Laboratory, University of Illinois at Urbana–Champaign, Urbana, IL 61801, USA}
\author {Dipanjan Chaudhuri}
\affiliation{Department of Physics and Materials Research Laboratory, University of Illinois at Urbana–Champaign, Urbana, IL 61801, USA}
\author {Farzaneh Hoveyda}
\affiliation{Department of Physics and Materials Research Laboratory, University of Illinois at Urbana–Champaign, Urbana, IL 61801, USA}
\author {Ali Husain}
\affiliation{Department of Physics and Materials Research Laboratory, University of Illinois at Urbana–Champaign, Urbana, IL 61801, USA}
\author {John Schneeloch}
\affiliation{Condensed Matter Physics and Materials Science Department, Brookhaven National Laboratory, Upton, New York 11973, USA}
\author {Genda Gu}
\affiliation{Condensed Matter Physics and Materials Science Department, Brookhaven National Laboratory, Upton, New York 11973, USA}
\author {Philip Phillips}
\affiliation{Department of Physics and Materials Research Laboratory, University of Illinois at Urbana–Champaign, Urbana, IL 61801, USA}
\author {Bruno Uchoa}
\affiliation{Department of Physics and Astronomy, University of Oklahoma, Norman, OK 73069, USA}
\author {Tai-Chang Chiang}
\affiliation{Department of Physics and Materials Research Laboratory, University of Illinois at Urbana–Champaign, Urbana, IL 61801, USA}
\author {Peter Abbamonte}\email{abbamonte@mrl.illinois.edu}
\affiliation{Department of Physics and Materials Research Laboratory, University of Illinois at Urbana–Champaign, Urbana, IL 61801, USA}


\date{\today}% It is always \today, today,

\begin{abstract}

The density fluctuation spectrum
captures many fundamental properties of strange metals. Using momentum-resolved electron energy-loss spectroscopy (M-EELS), we recently showed that the density response of the strange metal Bi$_2$Sr$_2$CaCu$_2$O$_{8+x}$ (Bi-2212) at large momentum, $q$, exhibits a constant-in-frequency continuum [Mitrano, PNAS \textbf{115}, 5392 (2018); Husain, PRX \textbf{9}, 041062 (2019)] reminiscent of the marginal Fermi liquid (MFL) hypothesis of the late 1980s [Varma, PRL \textbf{63}, 1996 (1989)]. However, reconciling this observation with infrared (IR) optics experiments, which show a well-defined plasmon excitation at $q \sim 0$, has been challenging. Here we report M-EELS measurements of Bi-2212 using 4$\times$ improved momentum resolution, allowing us to reach the optical limit. 
For momenta $q<0.04$ r.l.u., the M-EELS data show a plasmon feature that is quantitatively consistent with IR optics. For $q>0.04$ r.l.u., the spectra become incoherent with an MFL-like, constant-in-frequency form. We speculate that, at finite frequency, $\omega$, and nonzero $q$, some attribute of this Planckian metal randomizes the probe electron, causing it to lose information about its own momentum. 

%\begin{description}
%\end{description}
\end{abstract}


\maketitle

%\tableofcontents
%\section{Introduction}
\section{\label{sec:level1}Introduction }
The strange metal is a peculiar phase of matter whose resistivity violates the Mott–Ioffe–Regel limit \cite{martin1990normal,takagi1992systematic} and exhibits Planckian dissipation, conjectured to represent a quantum limit on dissipation in a many-body system \cite{zaanen2019planckian,hartnoll2015theory,bruin2013similarity,legros2019universal}. There is currently no generally accepted paradigm for understanding  strange metals, with approaches varying from quantum criticality to holographic duality \cite{zaanen2019planckian,phillips2022stranger}. 

A useful early framework for describing strange metals is the so-called marginal Fermi liquid (MFL) phenomenology, which hypothesizes that the polarizability, $\Pi''(q,\omega)$, is proportional to $1/T$ for $\omega<T$, constant for $\omega>T$, and independent of momentum, $q$ \cite{varma1989phenomenology,littlewood1991phenomenology}. This hypothesis explains many defining properties of strange metals, including the linear-in-$T$ resistivity \cite{takagi1992systematic}, quasiparticle lifetime $\sim(\omega^2+T^2)^{1/2}\;$ \cite{,Vishik2010,reber2019unified}, and a frequency-dependent conductivity that exhibits a power law at optical frequencies,  $\sigma(\omega)\sim\omega^{-2/3}\;$ \cite{levallois2016temperature,marel2003quantum}, which implies a renormalized scattering rate, $1/\tau^*\sim\omega\;\,$  \cite{marel2003quantum}. 

Despite its success, the MFL hypothesis seems unphysical. Any quasiparticle-based framework, for example based on the random phase approximation (RPA), would result in a polarizability that is highly $q$ dependent \cite{mitrano2018anomalous}. Measurements of the density response of strange metals at nonzero $q$ are therefore greatly needed. 

We previously performed momentum-resolved electron energy-loss spectroscopy (M-EELS) measurements, in reflection geometry, of optimally doped Bi$_2$Sr$_2$CaCu$_2$O$_{8+x}$ (Bi-2212), a strange metal, at large $q$. We observed a continuum that, for all $q>0.05$ reciprocal lattice units (r.l.u.), is constant in frequency for $\omega > 0.1$ eV \cite{mitrano2018anomalous,husain2019crossover}, reminiscent of the MFL hypothesis. For $q\sim0.05$ r.l.u., we observed a plasmon, consistent with early reflection EELS at this momentum \cite{schulte2002interplay}. However, its lineshape is significantly broader than that observed in infrared (IR) optics experiments at $q \sim 0$ \cite{levallois2016temperature}. This raises the question of how M-EELS data at nonzero $q$ relate to optical data in the $q\to 0$ limit \cite{levallois2016temperature,marel2003quantum,basov2005electrodynamics}. 

Quantitatively comparing M-EELS and IR optics is not trivial because they measure different charge response functions. Optics measures the complex dielectric function, $\epsilon(q,\omega) = [1+V(q) \chi(q,\omega)]^{-1}$, where $V(q)=e^2/\epsilon_0 q^2$ and $\chi(q,\omega)$ is the charge response function of the system, in the small momentum ($q\rightarrow 0$) limit. By contrast, M-EELS measures the dynamic charge susceptibility of the surface of a material (called the ``surface response") \cite{evans1972theory,mills1975scattering,ibach2013electron,vig2017measurement}, its doubly differential scattering cross section being given by,

\begin{equation}
        \frac{\partial^2\sigma}{\partial\Omega\partial \omega} = \sigma_0
        N(\omega)
        V_\mathrm{eff}^2(\textbf{q})
        \chi''_s(q,\omega),
\end{equation}

\noindent where $N(\omega) = \left [ \pi (e^{-\hbar \omega/k_B T}-1) \right ]^{-1}$ is the Bose factor,

\begin{equation}
    V_\mathrm{eff}(\textbf{q})=\frac{e^2/\epsilon_0}
    {q^2+(k_z^i+k_z^s)^2}
    \label{mtx_ele}
\end{equation}

\noindent is the Coulomb matrix element, $k_z^i$ and $k_z^s$ are the out-of-plane components of the momenta for the incoming and scattered electrons respectively, $\sigma_0$ is a constant defined in Ref. \cite{vig2017measurement}, and 

\begin{equation}
    \chi''_s(q,\omega)=
    \int_0^\infty dz_1 dz_2
    e^{-q(z_1+z_2)}
    \chi''(q,\omega,z_1,z_2)
    \label{eels_chi}
\end{equation}

\noindent is the response function for reflection EELS measurements, and is sometimes called the ``surface loss function" \cite{evans1972theory,vig2017measurement}. $\chi(q,\omega,z_1,z_2)$ is the charge response of a semi-infinite system, $q$ representing the in-plane momentum, and $z_1$ and $z_2$ representing the depth below the surface \cite{evans1972theory,vig2017measurement}. M-EELS and IR optics are therefore closely related, but a direct comparison requires some subtle analysis. 

Further, IR optics probes the material at very small $q$, $\sim 0$, while M-EELS measurements have previously focused on large $q$ \cite{mitrano2018anomalous,husain2019crossover}. Therefore, a quantitative comparison also requires M-EELS measurements to be performed at sufficiently small $q$ such that the optical limit is reached. In this limit, effects from the Coulomb matrix element (Eq. (\ref{mtx_ele})) are important, and it's also crucial to properly account for the finite momentum resolution of the measurement. 

In this study, we account for all these effects to investigate the consistency between M-EELS and optics and to improve the overall understanding of the density response of a strange metal at all values of $q$. 

%\section{Experiment}
\section{\label{sec:level1}Experiment}
\begin{figure}
    \centering
    \includegraphics[width=0.45\textwidth]{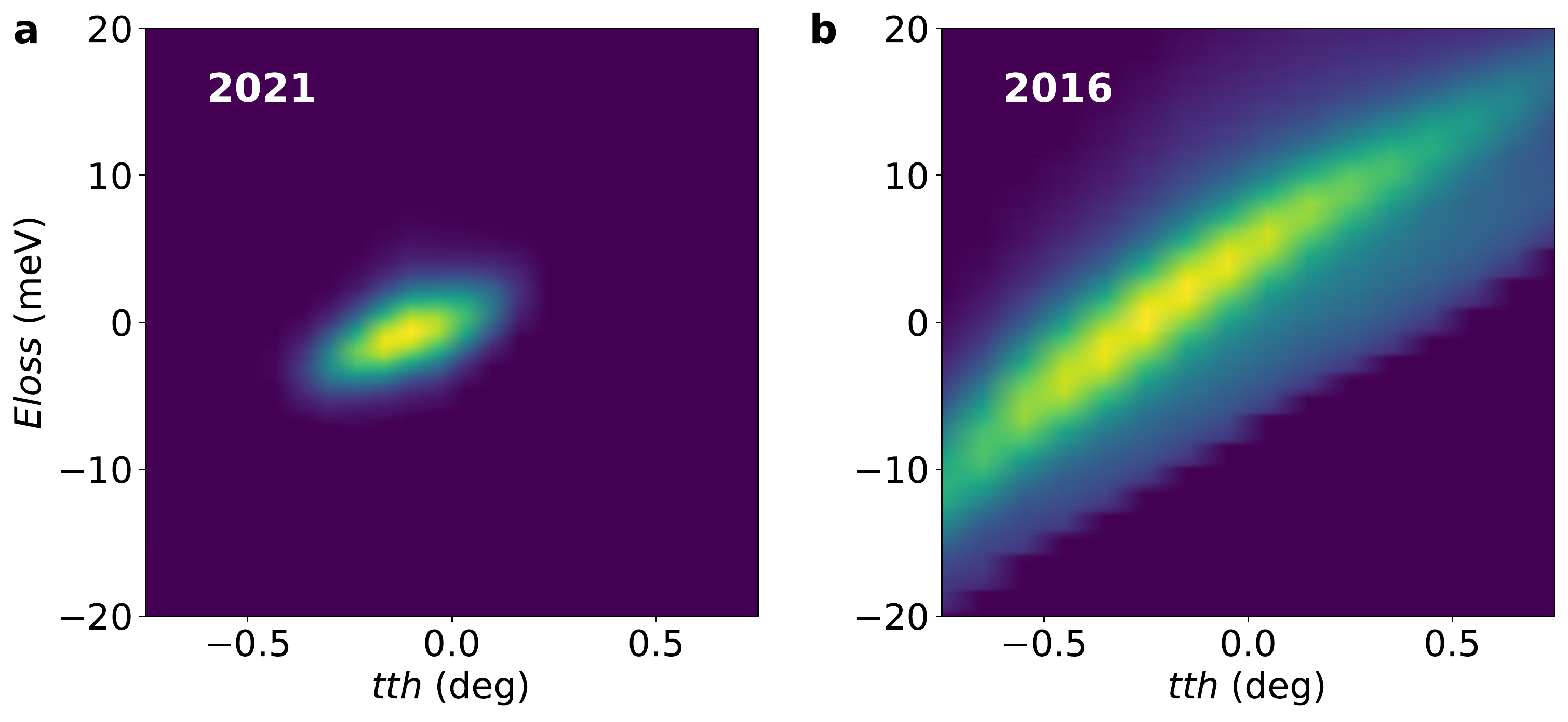}
    \caption{Phase space area of the incident electron beam in our M-EELS instrument, measured by performing an energy-angle map using the electron analyzer for (a) the current measurements and (b) for the measurements presented in Refs. \cite{mitrano2018anomalous,husain2019crossover}}
    \label{Beam_resl}
\end{figure}

Optimally doped single crystals of strange-metal-phase Bi-2212, with superconducting transition temperature, $T_c= 91$ K, were grown using the floating zone method described previously \cite{wen2008}. While the oxygen stoichometry, $x$, is not known precisely, this $T_c$ value correponds to a doping concentration of $p=0.16$ \cite{presland1991general}. 

Surface reflection M-EELS measurements of optimally doped Bi-2212 were done with an Ibach type, HR-EELS spectrometer \cite{ibach2013electron} using a beam energy of 50 eV and energy resolution $\Delta E=5$ meV. 
Noting that the M-EELS data  near 1 eV energy loss show no temperature dependence for optimal doping \cite{mitrano2018anomalous,husain2019crossover}, we performed reflection M-EELS measurements at room temperature, $T=300$ K. 
High momentum accuracy was achieved by motorizing the scattering angle, called ``tth," and aligning the axis of rotation to the center of a eucentric sample goniometer with a sphere of confusion of $\sim 80$ microns. Bi-2212 single crystals were cleaved {\it in situ}, and the orientation matrix determined by locating the (0,0,20) specular rod and (1,0,20) Bragg reflections. Momenta in this article are expressed using Miller indices in reciprocal lattice units, i.e., ($h,k,l$) represents $\textbf{q}=2\pi(h/a,k/b,l/c)$, where for Bi-2212 in tetragonal units $a=b=3.81${\AA}, $c=30.8${\AA}. 
Sample and detector angles were rotated in coordination with the analyzer pass voltage to keep both the in-plane and out-of-plane momentum transfer fixed during loss scans ($l=20$ for all measurements).

The current measurements were done with 
improved beam tuning techniques that optimize the full phase space area of the beam, as illustrated in Fig. \ref{Beam_resl}. This resulted in an energy resolution of 5.6 meV and a momentum resolution of 0.02 $\AA^{-1}$, which is $\sim 4\times$ better than our previous studies \cite{mitrano2018anomalous,husain2019crossover} (Note that while the beam resolution stated in Refs. \cite{mitrano2018anomalous,husain2019crossover} was 4 meV, this referred to a vertical cross section through the beam profile in Fig. \ref{Beam_resl}. The integrated resolution was, in actuality, closer to 24 meV).

%\sout{We denote momenta using Miller indices in reciprocal lattice units, i.e., ($H,K,L$) represents $\textbf{q}=2\pi(H/a,K/b,L/c)$, where for Bi-2212 in tetragonal units $a=b=3.81${\AA}, $c=30.8${\AA}. Crystals were cleaved {\it in situ}, and the orientation matrix determined by locating the (0,0,20) specular rod and (1,0,20) Bragg peak. 
%Sample and detector angles were rotated in concert with the analyzer pass voltage to keep both the in-plane and out-of-plane momentum transfer fixed during loss scans ($L=20$ for all measurements).}
%\sout{The current measurements were done with $\sim 4\times$ better momentum resolution than our previous studies} \cite{mitrano2018anomalous,husain2019crossover}\sout{, achieved through improved tuning of the instrument (see Supplement and Fig. S1).}
%\textcolor{red}{, with an HR-EELS spectrometer outfitted with a multi-axis sample goniometer allowing both high momentum accuracy and resolution \cite{vig2017measurement} (see Supplement for more details about experiment setup).} 

\begin{figure}
    \centering
    \includegraphics[width=0.48\textwidth]{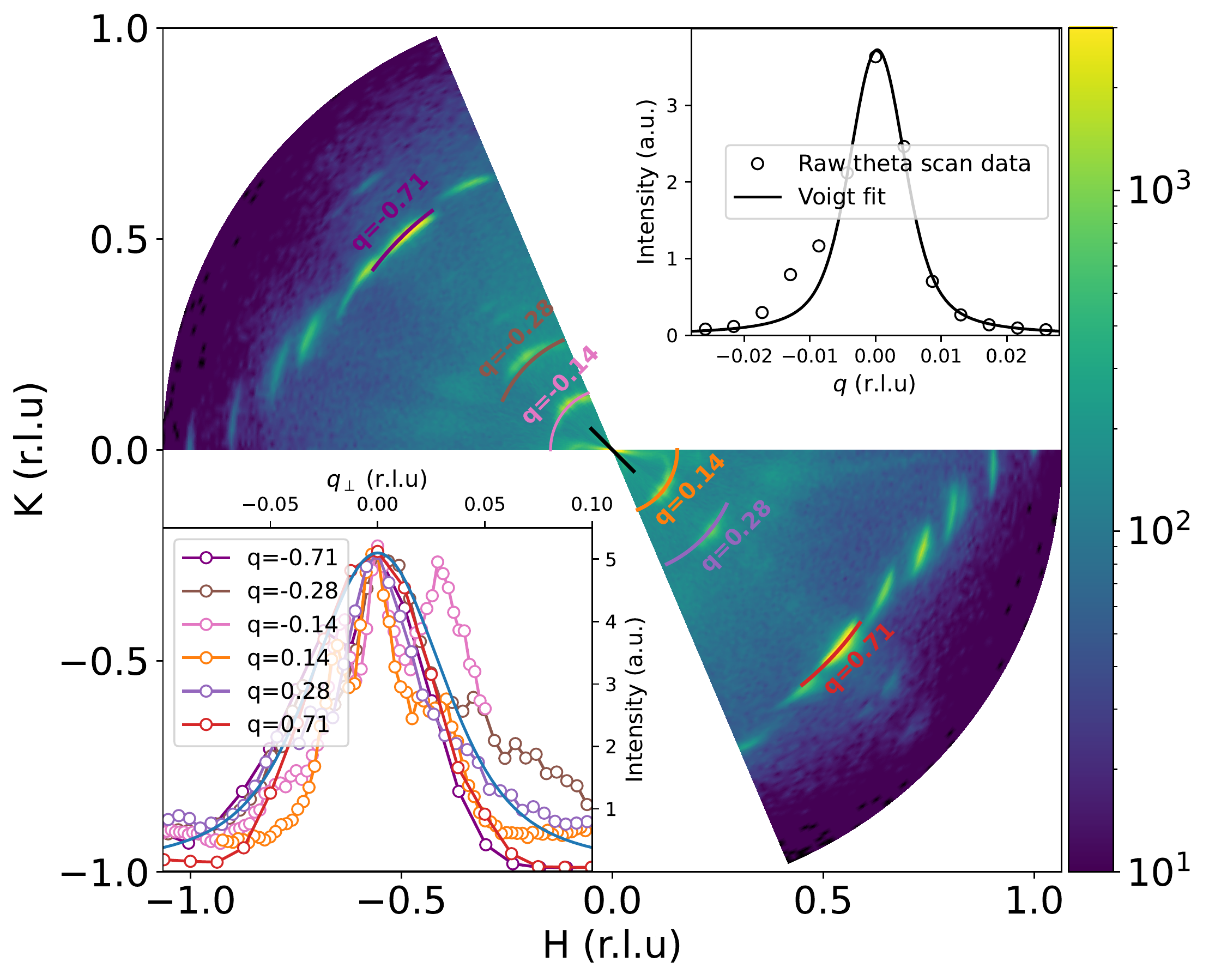}
    \caption{Elastic momentum map of the strange metal phase of Bi-2212 at room temperature. The bright spots are elastic Bragg reflections, most of which come from the well-known supermodulation \cite{damascelli2003}. (lower inset) $\phi$ scans through a selection of reflections with different values of $q$.
    The structure in some of the curves in the inset arises from imperfect periodicity of the structural supermodulation\cite{damascelli2003}. Here, $q_\perp$ represents the component of the momentum perpendicular to both $q$ and $q_z$. The blue line represents a Voigt fit to determine the out-of-plane momentum resolution, $\Delta q_V$.
    (upper inset) $\theta$ scan through the specular reflection at $q=0$ (black points). The Voigt fit (black line) gives the momentum resolution in the scattering plane, $\Delta q_H$.}
    \label{Thetaphi}
\end{figure}

Quantitatively comparing M-EELS to IR optics requires precise knowledge of the momentum resolution, which is determined by the divergence angle of the incident beam ({\it viz.} Fig. \ref{Beam_resl}), the angular acceptance of the analyzer \cite{EresNote}, and the quality (flatness and roughness) of the sample surface. 

We quantified our momentum resolution by performing a broad, elastic scattering ($\omega=0$) map of the cleaved Bi-2212 surface (Fig. \ref{Thetaphi}). This diffraction pattern shows the (1,0) and (-1,0) Bragg reflections, and numerous reflections from the well-known Bi-2212 structural supermodulation  \cite{damascelli2003}. The widths of these reflections give the momentum resolution convolving beam and surface effects. Fitting the curves with Voigt functions, we obtain momentum resolutions $\Delta q_H = 0.01013(41)$ r.l.u.
in the scattering plane and
$\Delta q_V=0.0744(29)$ r.l.u. in the out-of-plane direction (see Appendix \ref{Momentum resolution}).

%The most direct way to quantify the momentum resolution is to measure, {\it in situ} after cleaving, the width of Bragg reflections from the sample surface, which are resolution-limited and convolve both beam and surface quality effects. 
%We performed a broad, elastic scattering ($\omega=0$) map, Fig. \ref{Thetaphi}, which shows the (1,0) and (-1,0) Bragg reflections, and many reflections from the well-known structural supermodulation in this material \cite{damascelli2003}.
%As explained in Supplement, \sout{we measured the angular width of several Bragg peaks in both the $\theta$ and $\phi$, directions, which represent rotations in the scattering plane and azimuthal rotations around the surface normal, respectively (Fig. \ref{Thetaphi}, insets).} Fitting the curves with Voigt functions, we find $\Delta q_H = 0.01013(41)$ r.l.u.
%in the scattering plane and
%$\Delta q_V=0.0744(29)$ r.l.u. in the out-of-plane direction.

\section{\label{sec:level1}Results}
\begin{figure}
    \centering
    \includegraphics[width=0.45\textwidth]{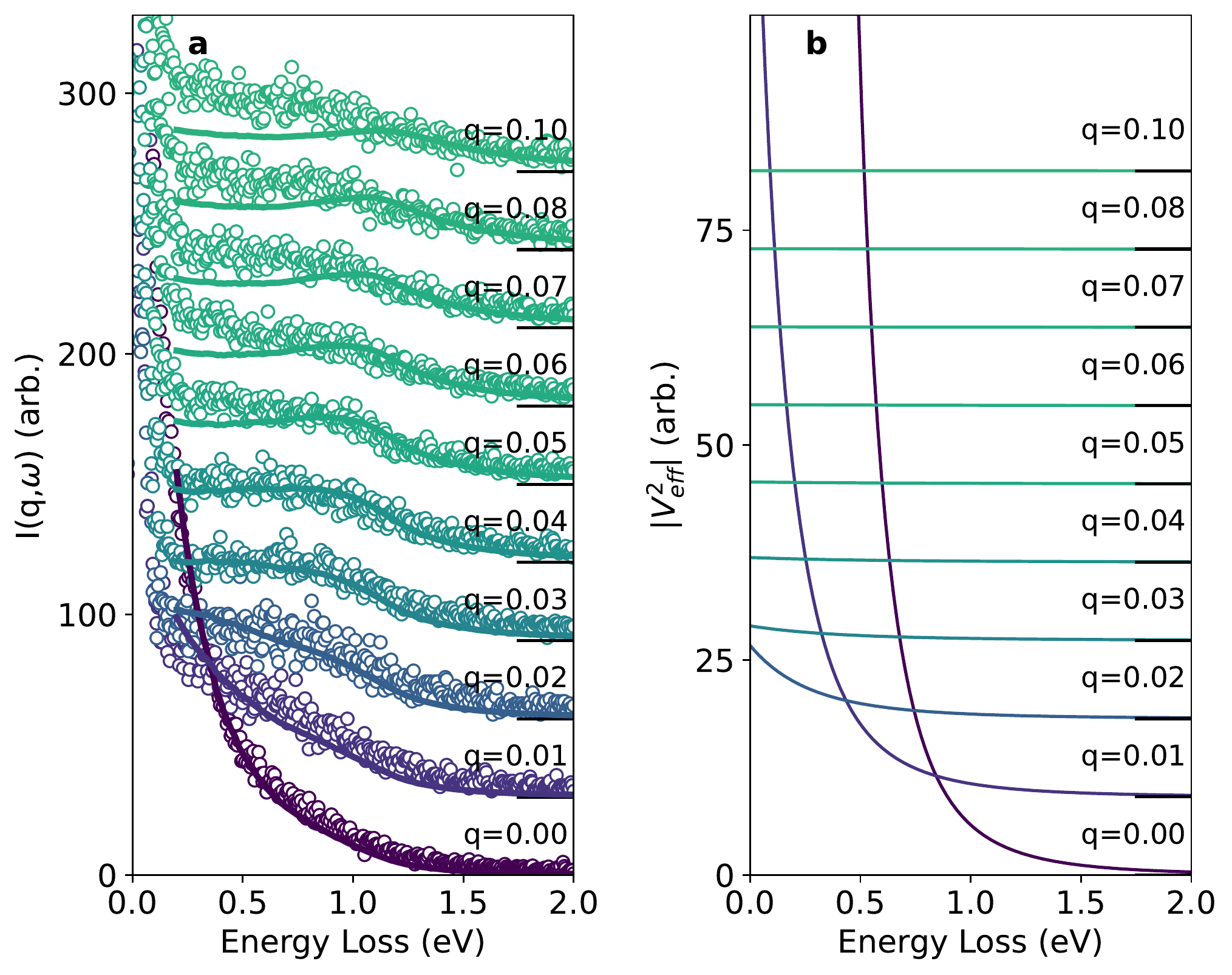}
    \caption{(a) Raw M-EELS data at different $q$ (open circles), scaled to constant spectral weight and offset for plotting purposes. Solid lines are the calculated M-EELS intensity from Eq. (\ref{total_chi}) (see text). (b) Matrix element $V_{eff}^2$ for each of the momenta showed in panel (a). These curves are also offset for clarity.} 
    \label{Rawdata}
\end{figure}

M-EELS spectra from Bi-2212 at $T=300$ K are shown in Fig. \ref{Rawdata}(a). Here, $q = \sqrt{h^2 + k^2}$ is in-plane momentum transfer in the (1,-1) direction in reciprocal lattice units (r.l.u.) \cite{mitrano2018anomalous,husain2019crossover}, $h$ and $k$ being Miller indices, and is held fixed during energy scans. The out of plane momentum, $l = 20$ r.l.u., for all measurements, and we focus on the long-wavelength limit, with $q \leq 0.1$ r.l.u. (Note that $l=20$ represents the momentum transferred to the probe electron, which describes the geometry of the experiment, and does not mean we are creating excitations only with momentum $q_z = 2 \pi l/c$, since $q_z$ is not perfectly conserved in a surface measurement. See Sections IV and V). 
For plotting, the data were scaled to constant spectral weight, defined as the first frequency moment of the loss spectrum. Otherwise, these are the raw data, not corrected for matrix elements, finite resolution, or interference from the elastic line. 

For $q>0.06$ r.l.u., the spectra show the continuum reported previously  \cite{mitrano2018anomalous,husain2019crossover}. For $0.02 < q < 0.04$ r.l.u., the spectra exhibit a bump-like structure at $\sim$ 1 eV arising from the 1 eV plasmon \cite{levallois2016temperature}, as demonstrated previously \cite{schulte2002interplay,vig2017measurement,mitrano2018anomalous}. For $q<0.02$ r.l.u., the spectra show a strongly energy-dependent tail, which is a rapidly decreasing function of $\omega$. 

%\textcolor{red}{[PICK IT UP HERE]}

This tail is an effect of the Coulomb matrix element, $V_\mathrm{eff}$ (Eq. (\ref{mtx_ele})) \cite{li2022geometric}. Fig. \ref{Rawdata}(b) plots $V_\mathrm{eff}^2$ against $\omega$ for the same momenta as in Fig. \ref{Rawdata}(a). For $q \gtrsim 0.03$ r.l.u., $V_\mathrm{eff}^2$ is basically constant. However, for $q \lesssim 0.03$ r.l.u., $V_\mathrm{eff}^2$ is a rapidly decreasing function of $\omega$, since the values $k_z^i$ and $k_z^s$ change during an energy scan when $q$ and $q_z$ are held fixed. The 1 eV ``bump" and continuum in Fig \ref{Rawdata}(a), not visible in Fig. \ref{Rawdata}(b), are properties of Bi-2212 that we aim to compare to IR optics data.

\section{\label{sec:level1}Density response of a layered metal}
We now establish a quantitative relationship between M-EELS and IR optics measurements. 
Bi-2212 is a layered, quasi-2D material in which interlayer hopping is significantly smaller than hopping within the layers \cite{damascelli2003}. The problem of the density response of a system of weakly coupled, metallic layers
was solved analytically by Jain and Allen in 1985 \cite{jain1985dielectric}. 
In their analysis, they considered a semi-infinite, layered system, with a well-defined surface, in which the individual metallic layers are arbitrarily thin and coupled only by the Coulomb interaction. The interlayer tunneling between the layers was assumed to be negligible. Their analysis was intended to interpret Raman scattering experiments from semiconductor superlattices \cite{Olego1982}, but their derivation of the density response function is general and may be applied just as well to M-EELS data from Bi-2212. 

Solving the Dyson equation, Jain and Allen found that the density response of a semi-infinite stack of metallic layers consists of two distinct terms corresponding to the bulk and surface responses of the material\cite{jain1985dielectric}, 

\begin{equation}
\begin{aligned}     
    \chi^\mathrm{bulk}_{l,l'}=\Pi^0\left[\delta_{l,l'}+\Pi^0 V\sinh(qd)(b^2-1)^{-\frac{1}{2}}u^{-|l-l'|}\right]
    \label{bulk_chi}
\end{aligned}
\end{equation}
\begin{equation}
\begin{aligned}
    \chi^\mathrm{surface}_{l,l'}=(\Pi^0)^2V\frac{u^2A-2uB+C}{2u^2(b^2-1)Q}u^{-(l+l')}
    \label{surface_chi}
\end{aligned}
\end{equation}

\noindent where $l$ and $l'$ are layer indices. $V=V(q)$ represents the in-plane, 2D Fourier transform of the Coulomb interaction. $b$, $u$, $A$, $B$, $C$ and $Q$ are complex functions of the polarizability, $\Pi^0$, the in-plane momentum, $q$, and the layer spacing, $d$, that Jain and Allen defined in their paper\cite{jain1985dielectric}.

The quantity $\Pi^0=\Pi^0(q,\omega)$ is the polarizability of a single layer, which contains all the microscopic physics of the CuO$_2$ planes. The assumption of discrete layers is valid as long as the thickness of a CuO$_2$ bilayer $\ll1/q$. This constraint is satisfied for our smallest momenta, $q \sim 0.01$ r.l.u., though not for the largest momentum, $q=0.1$ r.l.u. \cite{bilayerNote}. We will return to this point later in our discussion. 

The difference between the bulk $\chi^\mathrm{bulk}_{l,l'}$ and the surface $\chi^\mathrm{surface}_{l,l'}$ is that the former depends only on the distance between the layers, $l-l'$, and therefore has the same translational symmetry as the bulk of the material, while the latter decays exponentially with $l+l'$, so is localized near the surface. 

While Eqs. (\ref{bulk_chi}-\ref{surface_chi}) were developed for analyzing Raman scattering data, we can use them to express the the M-EELS response (Eq. (\ref{eels_chi})) in terms of the layer polarizability,

\begin{equation}
    \chi_s = \sum_{l,l'} \left ( \chi^\mathrm{bulk}_{l,l'} + \chi^\mathrm{surface}_{l,l'} \right ) 
    e^{-q l d} e^{-q l' d}
    \label{total_chi}
\end{equation}

\noindent where $d = c/2$ is the bilayer spacing. The only unknown in Eq. (\ref{total_chi}) is the polarizability of a single layer, $\Pi^0(q,\omega)$. At zero momentum, $q=0$, $\Pi^0$ is directly related to the dielectric function measured with IR optics \cite{PinesNozieres1973},

%\begin{equation}
%    \epsilon(\omega) = \epsilon_\infty 
%    - V(q) \, \Pi^0(q,\omega)
%\end{equation}

\begin{equation}
    \Pi^0(q,\omega) \Bigr |_{q \sim 0}
    =\frac{\epsilon_{\infty}-\epsilon(\omega)}{V_\mathrm{3D}} d
    \label{Eq_D0}
\end{equation}

\noindent where $\epsilon_\infty$ is the background dielectric constant and $V_\mathrm{3D}(q)$ is the three-dimensional Fourier transform of the Coulomb interaction.

Hence, we have found a way to make an explicit and quantitative comparison between M-EELS and IR optics experiments. Using Eq. (\ref{Eq_D0}), we can determine the polarizability of a single layer directly from the optical data, at least in the small momentum limit, $q \sim 0$. This function can then be used to evaluate Eq. (\ref{total_chi}), and compared directly to the M-EELS data without any adjustable parameters. 

The only limitation of this approach is that IR optics gives us the value of $\Pi^0(q,\omega)$ only at $q=0$. 
Because optics experiments only probe the long wavelength response, comparison at larger values of $q$ is not possible. 
But this can serve the purpose of validating the M-EELS data at small $q$, which will increase confidence in measurements at larger momenta that optical techniques cannot reach. 

We emphasize here that the current analysis, and our use of the Jain-Allen framework, is not a microscopic theory of the marginal Fermi liquid or of Bi-2212. We offer no microscopic explanation for the form of $\Pi^0(q,\omega)$. Our analysis is just a way of validating M-EELS data against much better established IR measurements in the long-wavelength limit, where the two techniques should be equivalent. 

\begin{figure}
    \centering
    \includegraphics[width=0.48\textwidth]{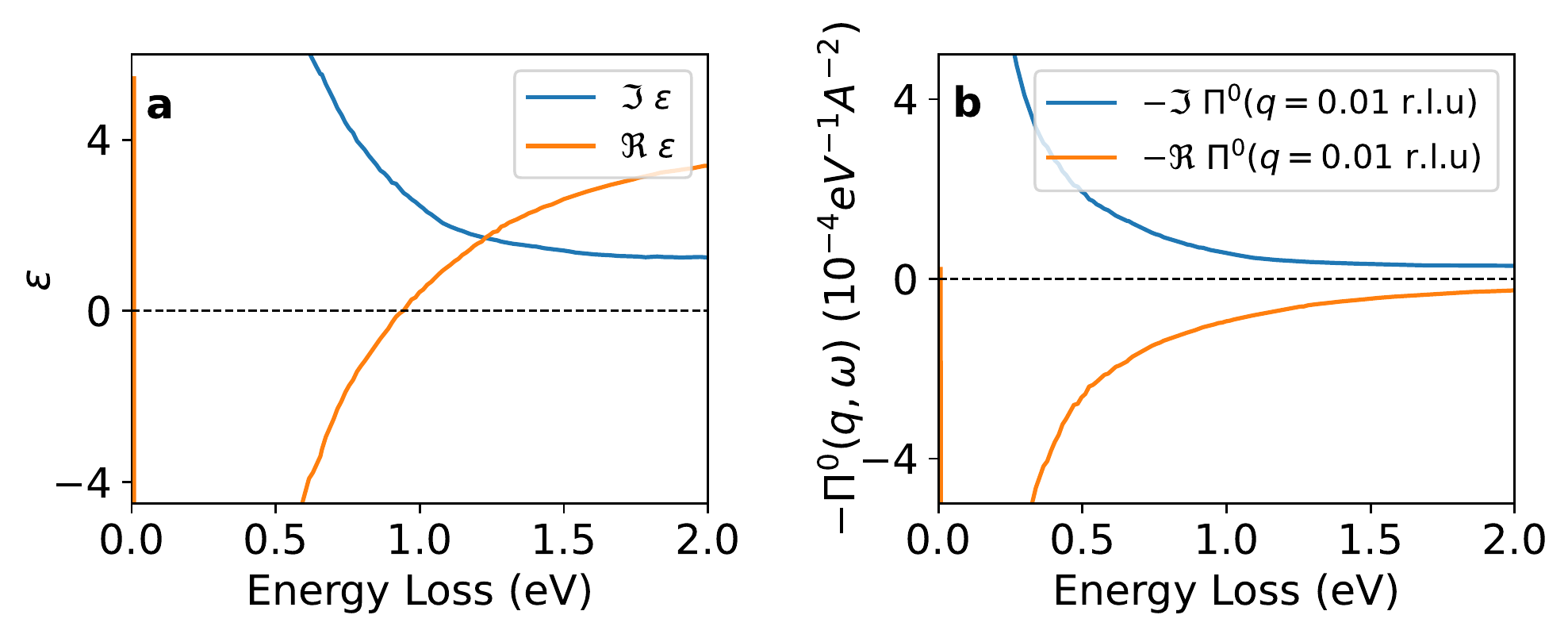}
    \caption{(a) Dielectric function for optimally doped Bi-2212 from Refs.\cite{marel2003quantum,levallois2016temperature}, showing a zero crossing of the real part at $\omega \sim$ 1 eV. (b) Calculated 2D polarizability, for $q=0.01$ r.l.u. from (a), using Eq. (\ref{Eq_D0}).}
    \label{D0}
\end{figure}

%\begin{figure*}[t!!!]
%    \centering
%    \includegraphics[width=0.48\textwidth]{Expected_D.png}
 %   \caption[Caption for D]{Predicted M-EELS response functions for different $q$, calculated from IR optics data (Fig. 3). (a) Bulk contribution. (b) Surface contribution. (c) Total contribution. The red and blue arrows in panel (a) indicate the energy of the out-of-phase ($q_z=\pi/d$) and in-phase ($q_z=0$) parts of the plasmon continuum, respectively, at each in plane-momentum $q$.}
%    \label{Expected_D}
%\end{figure*}

\section{\label{sec:level1}Relating M-EELS to IR Optics}

Fig. \ref{D0} shows $\Pi^0(q,\omega)$ for Bi-2212 determined from the IR optical data of Refs. \cite{levallois2016temperature,marel2003quantum} at $T=300$ K, using their value $\epsilon_{\infty}=4.5$. 
The dielectric function, $\epsilon(\omega)$ (Fig. \ref{D0}(a)), exhibits a non-Drude response with conductivity $\sigma(\omega) \sim \omega^{-2/3}\;$ \cite{marel2003quantum}. The polarizability determined from Fig. \ref{D0}(a) using Eq. (\ref{Eq_D0}) is shown in Fig. \ref{D0}(b). Note that $\Pi^0(q,\omega)$ is formally zero at $q=0$, because $V(q)$ in Eq. (\ref{Eq_D0}) diverges as $q \rightarrow 0$. Thus Fig. \ref{D0}(b) is evaluated at a small but nonzero value of $q$. 

While the polarizability in Fig. \ref{D0}(b) is valid only at small $q$, it allows us to check the consistency between M-EELS and IR in the optical limit. At larger $q$, discrepancies are expected, both because $\Pi^0(q,\omega)$ may be $q$-dependent, and because the thin layer approximation, $qd \ll 1$, breaks down. 

\begin{figure}
    \centering
    \includegraphics[width=0.48\textwidth]{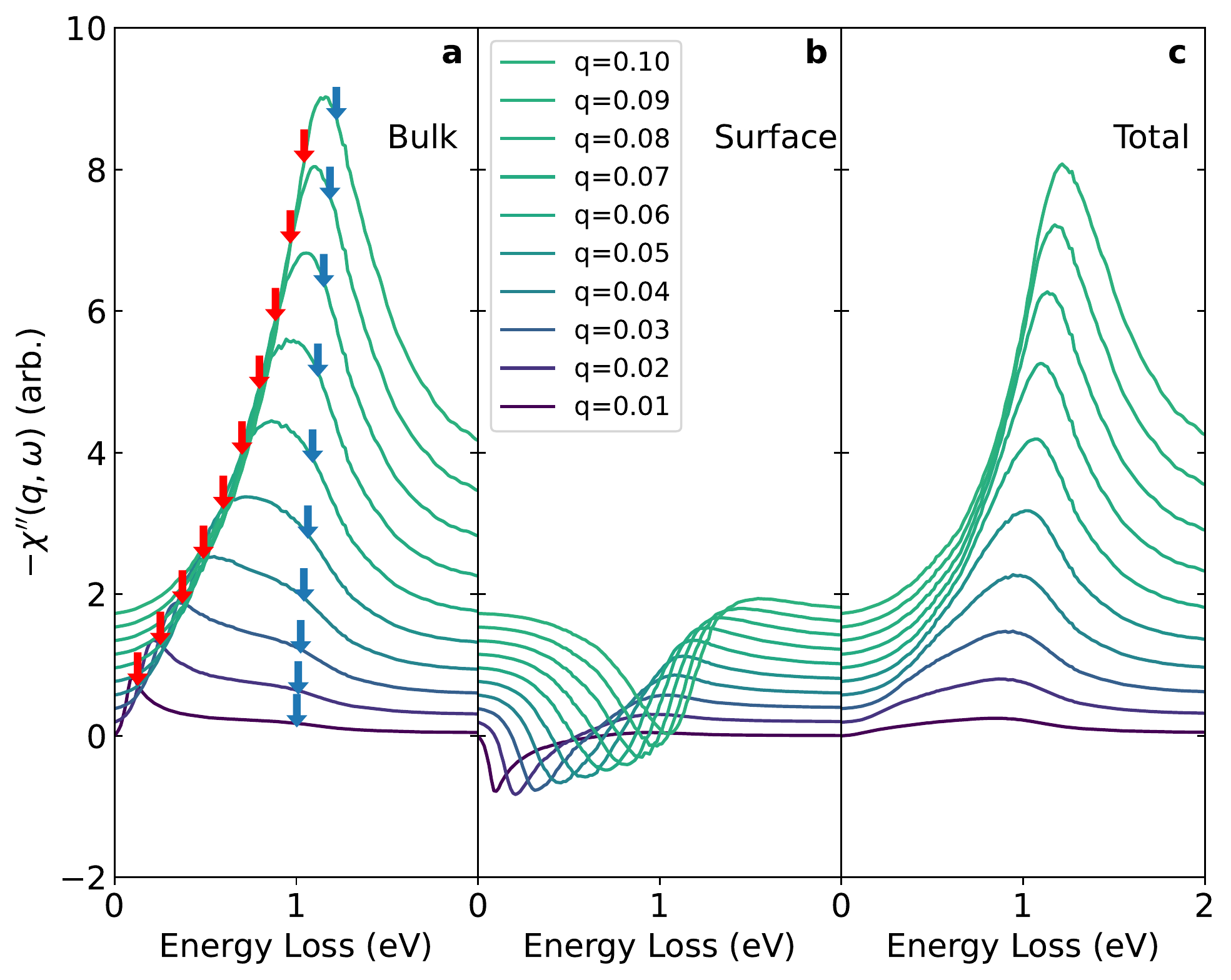}
    \caption[Caption for D]{M-EELS response functions for different $q$, predicted from the IR optics data of Fig. \ref{D0}. (a) Bulk response. (b) Surface response. (c) Total response. The red and blue arrows in panel (a) indicate the energies of the out-of-phase ($q_z=\pi/d$) and in-phase ($q_z=0$) parts of the plasmon continuum, respectively, at each in plane-momentum $q$, using $\omega_p(\textbf{q},q_z)=\left[\omega_{p,0}^2\frac{qd}{2}\frac{\sinh(qd)}{\cosh(qd)-\cos(q_zd)}\right]^{1/2}$ from the Fetter model \cite{fetter1974electrodynamics}.}
    \label{Expected_D}
\end{figure}

The bulk and surface response functions determined from Eqs. (\ref{bulk_chi}) and (\ref{surface_chi}), using $\Pi^0(q,\omega)$ in Fig. \ref{D0}, are shown in Fig. \ref{Expected_D}. No adjustable parameters were used; even the value of $\epsilon_\infty$ is known 
\cite{levallois2016temperature,marel2003quantum}.
The bulk response (Fig. \ref{Expected_D}(a)), at the lowest momentum $q=0.01$ r.l.u., shows a band of excitations beginning with a sharp peak at 98 meV, followed by a broad plateau that extends beyond the plasma frequency of 1 eV. This spectrum corresponds to the well-known spectrum of plasmons in a layered electron gas \cite{fetter1974electrodynamics}.
The broad band appears because M-EELS is a surface probe that does not conserve $q_z$ \cite{vig2017measurement}. The sharp, lowest energy feature corresponds to the gapless, out-of-phase mode at $q_z=\pi/d$ \cite{bozovic1990plasmons,fetter1974electrodynamics} reported in several recent RIXS experiments \cite{nag2020detection,hepting2018three,hepting2022gapped}. As $q$ is increased, the out-of-phase mode disperses in the expected way for small $q$. For sufficiently large q, the entire spectrum merges into a single, broad peak at 1.15 eV, which is similar to the frequency of the in-phase mode at $q_z=0$. 

Consider now the surface response, Fig. \ref{Expected_D}(b). This quantity can be negative so long as the total response (Eq. (\ref{total_chi})), the experimental observable, is positive. The same features are visible as in the bulk response, with an out-of-phase mode at low $q$ that merges with the rest of the spectrum at large $q$. 

The combined response is shown in Fig. \ref{Expected_D}(c). Curiously, adding the bulk and surface responses suppresses the out-of-phase ($q_z=\pi/d$) feature in the spectrum, an effect that was previously noted in Ref. \cite{jain1985dielectric}. The result is that most of the weight is concentrated near the in-phase $q_z=0$ plasma frequency. 

Note that the total response is predicted to be highly $q$-dependent, the plasmon being strongly dispersive, even though we assumed, by construction, that $\Pi^0$ is $q$-independent. 
This implies that the plasmon can acquire some dispersion purely from the Coulomb interaction between the layers. 

Finally, we compare the results in Fig. \ref{Expected_D}(c) with the experimental M-EELS data (Fig. \ref{Rawdata}(a)). We multiplied the full response (Fig. \ref{Expected_D}(c)) by the Bose factor and corresponding matrix elements (Fig. \ref{Rawdata}(b)), and then convolved with the momentum resolution determined from Fig. \ref{Thetaphi}. It was necessary to treat the overall magnitude as an adjustable parameter, since Figs. \ref{Expected_D}(c) and \ref{Rawdata}(a) have different units.  
The results are shown as solid lines in Fig. \ref{Rawdata}(a). 
At the smallest q values, $q<0.04$ r.l.u., the agreement is excellent, the calculated spectra reproducing both the rapidly decreasing tail at $q\leq0$, 0.01 r.l.u. and the plasmon feature at 1 eV.
Note that the shape of the calculated curve is not {\it exactly} the same as the experiment, falling slightly below in some regions of the spectrum and slightly above in others. But, overall, the main features of the data are reproduced.
We conclude that M-EELS and optics are quantitatively consistent in the $q\to0$ limit (in opposition to past conjectures \cite{phillips2022stranger}). 

At larger momenta, $q \gtrsim 0.04$ r.l.u., in the region of the MFL-like continuum \cite{mitrano2018anomalous,husain2019crossover}, the curves are no longer consistent. 
The Jain-Allen analysis predicts an asymmetric, peak-like feature that disperses to higher energy with increasing $q$. 
By contrast, the actual M-EELS spectra just show a momentum-independent continuum that does not visibly change for $q > 0.05$ r.l.u. \cite{mitrano2018anomalous,husain2019crossover}. 
As mentioned earlier, a discrepancy is expected, because $\Pi^0$ may be $q$-dependent, and because the thin-layer approximation, $qd \ll 1$, breaks down for $q \gtrsim 0.04$. We conclude that M-EELS and IR optics are quantitatively consistent in the $q \rightarrow 0$ limit, but not for appreciably large values of $q$.

\section{\label{sec:level1}Possibility of momentum scrambling}
\begin{figure}
    \centering
    \includegraphics[width=0.48\textwidth]{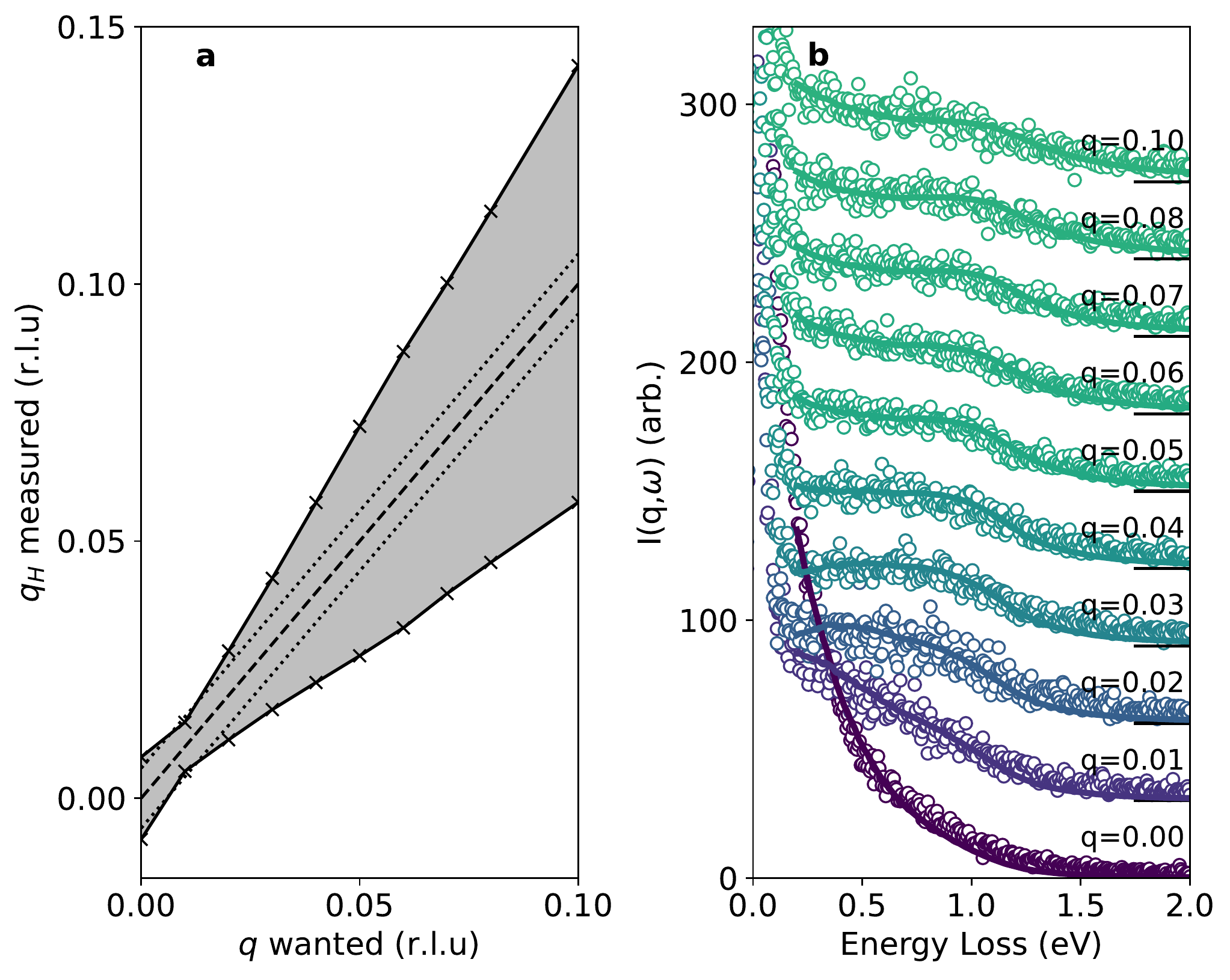}
    \caption{(a) Range of integrated momenta for the ``momentum scrambling" exercise described in Section VI. The horizontal axis represents the nominal, in-plane momentum. The dotted lines represents the experimental momentum resolution, $\Delta q_H$, determined from Fig. \ref{Thetaphi}. 
    The grey region represents the expanded integration range used to obtain the fits. (b) 
    Fits using the expanded momentum integration range (panel (a)) (solid lines) to the same experimental data shown in Fig. \ref{Rawdata}(a) (open circles). The fit at large $q$ is greatly improved compared to Fig. \ref{Rawdata}}
    \label{qScrambling}
\end{figure}

In the previous section, we treated the momentum resolutions, $\Delta q_H = 0.01013(41)$ r.l.u. and
$\Delta q_V=0.0744(29)$, 
as fixed properties of the instrument, which we quantified by fitting diffraction data. Good agreement between M-EELS and IR was achieved for small momenta $q<0.04$ r.l.u., where the two techniques must agree, by convolving the Jain-Allen susceptibility with these resolutions. 

In this section, we will treat the momentum resolution as  adjustable, to see if better agreement can be obtained for $q>0.04$ r.l.u.. We follow the same procedure outlined in Sections IV-V. However, we now treat the smaller of the two resolutions, $\Delta q_H$, as tunable, introducing an artificial Gaussian broadening in the in-plane direction (because $\Delta q_V$ is much larger, we keep this quantity fixed to the same value used earlier). There is no physical reason to expect the momentum resolution to be compromised in this way. So, for the moment, this should be considered a purely mathematical exercise.

%\textcolor{red}{Follow the same method in Fig. \ref{Thetaphi}, however, this time we use Gaussian function instead of Voigt function to simplify the computation since Voigt function has two adjustable parameters while Gaussian function just have one. This ends up with $\Delta q_H=0.0117(12)$ r.l.u., $\Delta q_V=0.0872(18)$ r.l.u. Then we treat $\Delta q_H$ as an adjustable parameter while keep $\Delta q_V$ fixed and fit the calculated M-EELS intensity with M-EELS raw data for each $q$ respectively. We only change $\Delta q_H$ because $\Delta q_H \ll \Delta q_V$ and small broadening in momentum width, if resulting purely from sample, should happen in the smaller one, i.e., $\Delta q_H$. }

The results of this tuning  are summarized in Fig. \ref{qScrambling}. Panel (a) shows the fit value of $\Delta q_H$ for different values of the in-plane momentum, $q$. Panel (b) shows the resulting fits compared to the experimental data, reproduced from Fig. \ref{Rawdata}. The calculation, which is still based on the same IR optical data, matches very well to the M-EELS data for all momenta, from $q=0$ all the way up to $q=0.1$ r.l.u.. The fit not only reproduces the rapidly decreasing tail at $q\leq$ 0.01 r.l.u., but also the plasmon feature at $q<$0.04 r.l.u. and the MFL-like continuum at $q\gtrsim$ 0.04 r.l.u. 

What this fit suggests, highly speculatively, is that some phenomenon is {\it randomizing the momentum of the probe electron}. As shown in Fig.~\ref{qScrambling}(a), the degree of randomization depends on the value of $q$, increasing for increasing in-plane momenta. For $q\leq$ 0.02 r.l.u., $\Delta q_H$ 
has the value estimated from elastic momentum map, meaning it is determined by instrument and sample surface effects (see Section II). As $q$ increases, however, $\Delta q_H$ increases monotonically such that $\Delta q_H/q \sim$ 0.88 is roughly a fixed number. 

It is hard to imagine what phenomenon could scramble the electron momentum in this way.  
Because the value of $\Delta q_H$ greatly exceeds the instrumental value, we speculate that this behavior originates from the material itself. Note that this behavior cannot be explained by multiple scattering, which tends to push spectral weight to higher energy by making multiple excitations \cite{warren1980theory,garciia2013multiple}. The broadening considered here is only in the momentum direction. We conjecture that this effect may be due to some previously unobserved property of Bi-2212.

%\section{Discussion \& conclusions}
\section{\label{sec:level1}Discussion}
Based on the above analysis, it is clear that, for momenta $q<0.04$ r.l.u., surface M-EELS data fully support the existence of a plasmon mode in the strange metal phase of Bi-2212, whose energy and lineshape are consistent with previous IR optics experiments \cite{marel2003quantum,levallois2016temperature}. At larger momenta, $q>0.04$ r.l.u., the data are no longer consistent with IR optics, and no longer support the existence of a well-defined plasmon, exhibiting instead a frequency-independent continuum reminiscent of the MFL hypothesis \cite{varma1989phenomenology,littlewood1991phenomenology}. This observation seems to contradict early, high-energy EELS measurements on Bi-2212, which observed a propagating, free-electron-like, RPA plasmon at all values of $q$ \cite{Nucker1989,wang1990electron}. However, more recent transmission EELS measurements with improved energy resolution, $<$20 meV, are inconsistent with early studies, and show a continuum feature similar to what is observed with M-EELS\cite{husain2019crossover,terauchi1995electron,terauchi1999development}. 
Further measurements are needed to reconcile these conflicting observations. 

There are several ways our Jain-Allen-based approach could have gone wrong. First, the polarizability, $\Pi^0(q,\omega)$, may strongly depend on $q$. It is tempting, then, to fit our data in Fig. \ref{Rawdata}(a) to determine the full, $q$-dependent $\Pi^0$ at $q \neq 0$. Doing so requires deconvolving the instrument resolution, which is highly statistically unstable, and would lead to the absurd conclusion that $\Pi^0$ exhibits a strong $q$ dependence that exactly cancels the $q$ dependence of the Jain-Allen formulas in such a way as to render the M-EELS data $q$-independent. Such a conclusion would be farfetched. 

The thin-layer approximation could also be breaking down. This problem could be corrected by accounting for the microscopic charge distribution within the CuO$_2$ layers, and would result in slightly different predictions. However, it seems unlikely to give a response that is frequency- and momentum-independent, like we observe.  

A final possibility is that some property of Bi-2212 randomizes the momentum of probe electrons undergoing losses of order $\omega_p$, causing the M-EELS scattering itself to become incoherent. Bi-2212 is a Planckian metal in which the scattering rate is thought to be near a quantum limit \cite{zaanen2019planckian,hartnoll2015theory,phillips2022stranger,patel2019theory}. Perhaps this effect influences the probe electrons, though only for momenta $q>0.04$ r.l.u.. It is worth considering whether the MFL effect itself may derive from momentum scrambling, i.e., translational symmetry being somehow dynamically broken, rather than an exotic spectrum of quantum critical fluctuations \cite{thornton2023jamming}.

\begin{acknowledgments}
We acknowledge helpful discussions with J\"org Fink and Dirk van der Marel. 
This work was primarily supported by the Center for Quantum Sensing and Quantum Materials, an Energy Frontier Research Center funded by the U.S. Department of Energy (DOE), Office of Science, Basic Energy Sciences (BES), under award DE-SC0021238. Growth of Bi-2212 single crystals was supported by DOE Grant No. DE-SC0012704. P.A. gratefully acknowledges additional support from the EPiQS program of the Gordon and Betty Moore Foundation, grant GBMF9452. B.U. acknowledges NSF grant DMR-2024864 for support.
\end{acknowledgments}

\appendix

\section{Momentum resolution}\label{Momentum resolution}
\begin{figure*}
    \centering
    \includegraphics[width=0.95\textwidth]{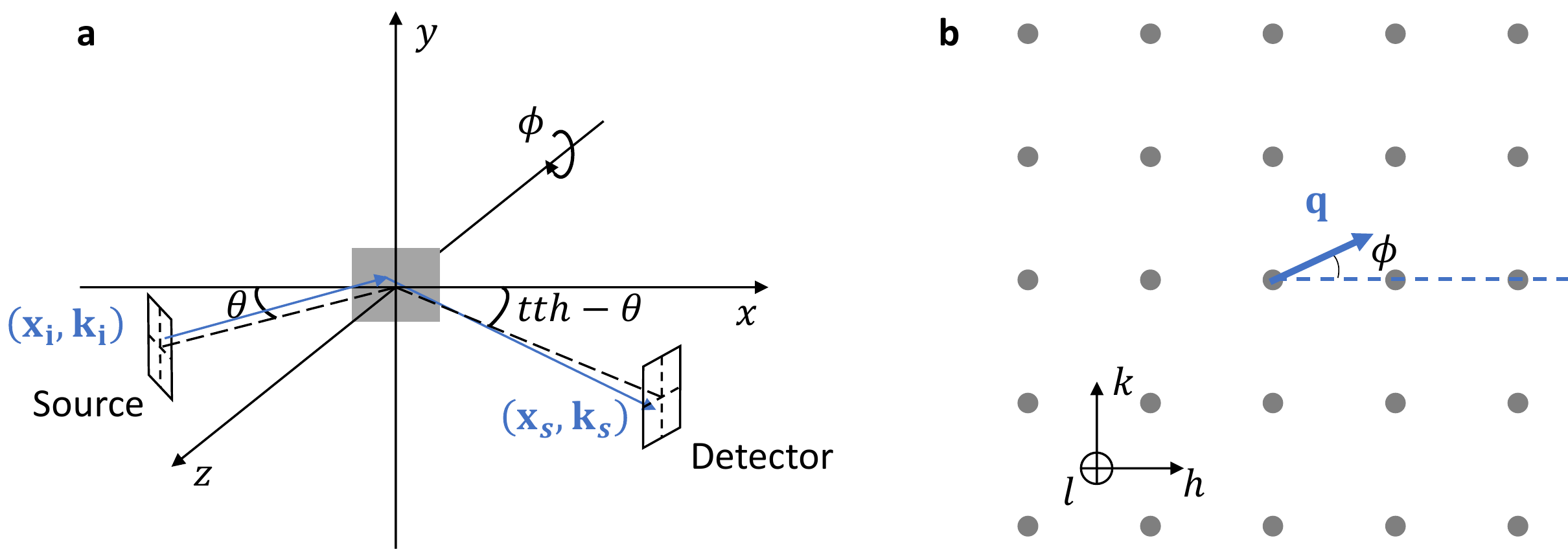}
    \caption{(a) Cartoon of M-EELS scattering process. Shaded rectangle is referred as sample. Blue arrows refer to incoming and scattered probing electrons. (b) As $\phi$ rotates, momentum transfer $\textbf{q}$ rotates accordingly in reciprocal lattice space. }
    \label{Cartoon_resl}
\end{figure*}
The current measurements were done with an improved energy and momentum resolution compared to Refs. \cite{mitrano2018anomalous,husain2019crossover}, allowing us to reach the optical limit where comparison with IR optics measurements is meaningful. Nevertheless, accurate comparison between M-EELS and IR measurements requires precise knowledge of the momentum resolution--not only in the horizontal scattering plane, but also in the vertical direction, perpendicular to the scattering plane. The resolution in both directions can be obtained by measuring the angular widths of the specular reflection as well as non-specular Bragg reflections from the Bi-2212 crystal itself (see Fig. \ref{Thetaphi}). 

The relationship between the widths of a Bragg reflection and the momentum resolution is illustrated in Fig. \ref{Cartoon_resl}. Here, the $x-z$ plane represents the horizontal, scattering plane, and the shaded rectangle represents the sample. 
$\theta$ corresponds to sample rotations around the $y$ direction, i.e., 
the angle of incidence on the sample surface. $\phi$ represents azimuthal sample rotations around its surface normal. If the momentum resolution were perfect, the momentum transfer would lie purely in the horizontal plane. Defined with respect to the sample surface (Fig. \ref{Cartoon_resl}(a)), the momentum transfer, $\textbf{q}=(q_x,q_y,q_z)$, where
\renewcommand\theequation{S\arabic{equation}}
\begin{equation}
\begin{cases}
    q_x=k_i\sin\theta-k_s\sin(tth-\theta)\\
    q_y=0\\
    q_z=k_i\cos\theta+k_s\cos(tth-\theta).
    \label{idea_surface}
\end{cases}
\end{equation}

\noindent 
Here, $(q_x,q_y)$ represents the component of the momentum in the plane of the sample surface, $q_x$ representing the horizontal irection, and $q_z$ is the component normal to the surface. 

The relevant quantity for scattering is is the momentum transfer in {\it sample} coordinates, $\overline{\textbf{q}} =(q_h,q_k,q_l)$, illustrated in Fig. \ref{Cartoon_resl}(b). If the azimuthal angle, $\phi=0$, then $\overline{\textbf{q}}=\textbf{q}$, i.e., $(q_x,q_y,q_z) = (q_h,q_k,q_l)$. As we rotate $\phi$, $\overline{\textbf{q}}$ rotates in sample coordinates, 

\renewcommand\theequation{S\arabic{equation}}
\begin{equation}
\begin{cases}
q_h=q_x\cos\phi\\
q_k=q_x\sin\phi\\
q_l=q_z.
\label{idea_sample}
\end{cases}
\end{equation}
 
Low-energy electrons typically undergo two types of elastic scattering: (1) specular reflection off the sample surface, and (2) Bragg reflection off the crystal lattice. If the crystal is perfect, i.e., both crystallographically and in terms of the flatness of the sample surface, specular scattering will only be visible when the in-plane $(q_x,q_y)=(0,0)$. Similarly, Bragg scattering will only be visible when the in-plane wave vector coincides with a reciprocal lattice vector of the sample surface, i.e., $(q_h,q_k) = 2 \pi (H,K)/a$, where $H$ and $K$ are integer Miller indices. Reflections of the latter type can be called LEED reflections. 
Under perfect conditions, both specular and Bragg reflections are infinitely sharp and exhibit zero angular width (though are broad in the $q_z$ direction). 

In a real scattering measurement, the beam is not perfectly collimated and will have a nonzero divergence in both the horizontal and vertical directions. Similarly, the crystal may have nonzero mosaicity, and the surface may be non-flat or have finite roughness. In this case, the momentum resolution of the measurement will be finite, and both specular and Bragg reflections will be broadeneed. The most direct way to measure the momentum resolution of a surface EELS measurement, then, is to measure the angular widths of various elastic reflections, {\it in situ} after the crystal is cleaved, which quantifies the momentum resolution incorporating both beam divergence and sample quality effects.

In this situation, Eqs. (\ref{idea_surface}) are no longer strictly true. For a given motor positions, $\theta$ and {\it tth}, the instrument will integrate over a range of momenta in the horizontal and vertical directions, 
\renewcommand\theequation{S\arabic{equation}}
\begin{equation}
\begin{cases}
    q_x=k_i\sin\theta-k_s\sin(tth-\theta)\pm\Delta q_H/2\\
    q_y=\pm\Delta q_V/2\\
    q_z=k_i\cos\theta+k_s\cos(tth-\theta)\pm\Delta q_z/2
    \label{real_surface}
\end{cases}
\end{equation}
\noindent where $\Delta q_H$ and $\Delta q_V$ represent the FWHM momentum resolutions in the horizontal and vertical directions. Because the scattering is broad anyway in $q_z$ direction, because the surface breaks translational symmetry in the $z$ direction, the effect of $\Delta q_z$ is unimportant and can be ignored. In terms of sample coordinates, Eqs. (\ref{real_surface}) have the form 
\renewcommand\theequation{S\arabic{equation}}
\begin{widetext}
\begin{equation}
\begin{cases}
    q_h=\left [ k_i\sin\theta-k_s\sin(tth-\theta) \right ]\cos\phi\pm (\Delta q_H/2)\cos\phi\mp (\Delta q_V/2)\sin\phi\\
    q_k=\left [ k_i\sin\theta-k_s\sin(tth-\theta) \right ]\sin\phi\pm (\Delta q_H/2) \sin\phi\pm (\Delta q_V/2) \cos\phi.
    \label{real_sample}
\end{cases}
\end{equation}
\end{widetext}

\noindent As a consequence of Eqs. (\ref{real_sample}), the Bragg condition $(q_h,q_k) = 2 \pi (H,K)/a$ is no longer rigid, and will be satisfied for a range of motor angles, $\theta$ and {\it tth}, defined by the momentum widths $\Delta q_H$ and $\Delta q_V$.

For case of elastic scattering, Eq. (\ref{real_surface}) gives a relationship between the FWHM momentum width of the specular reflection and the momentum transfer resolution in the horizontal direction, 

\begin{equation}
    \Delta q_x=\Delta q_H.
\end{equation}

\noindent Meanwhile, near a Bragg reflection $2 \pi (H,K)/a$, it is useful to define the momentum transfer perpendicular to the scattering plane, 
\begin{equation}
    q_\perp=\sqrt{q_h^2+q_k^2} \cdot \sin(\phi-\phi_0),
\end{equation}
where $\phi_0$ is the azimuthal angle of the Bragg peak of interest, $\phi_0\equiv\arctan{(K/H)}$. Using Eqs. (\ref{real_sample}), in a given diffractometer configuration, the experiment will integrate over a range of 
$q_\perp$ given by,
\begin{widetext}
\begin{equation}
    q_\perp=\left [ k_i\sin\theta-k_s\sin(tth-\theta) \right ]\sin(\phi-\phi_0)\pm (\Delta q_H/2) \sin(\phi-\phi_0)\pm (\Delta q_V/2) \cos(\phi-\phi_0)
    \label{q-perp}
\end{equation}
\end{widetext}

\noindent At Bragg condition, $\phi=\phi_0$, and Eq. (\ref{q-perp}) reduces to
\begin{equation}
    q_\perp=\pm\Delta q_V/2
\end{equation}
Eq. (\ref{q-perp}) implies that the momentum resolution of the measurement in the vertical direction, $\Delta q_V$, is simply the width of a Bragg peak in a scan that varies the value of $q_\perp$. An example of such a scan would be rotating the $\phi$ motor to rock the crystal through the Bragg condition, as shown in Fig. \ref{Thetaphi} of the main manuscript. The width of this scan directly gives 

\begin{equation}
    \Delta q_\perp=\Delta q_V.
\end{equation}

%\section{Reference}

\bibliography{jinplasmon}% Produces the bibliography via BibTeX.

\providecommand{\noopsort}[1]{}\providecommand{\singleletter}[1]{#1}%
\begin{thebibliography}{43}%
\makeatletter
\providecommand \@ifxundefined [1]{%
 \@ifx{#1\undefined}
}%
\providecommand \@ifnum [1]{%
 \ifnum #1\expandafter \@firstoftwo
 \else \expandafter \@secondoftwo
 \fi
}%
\providecommand \@ifx [1]{%
 \ifx #1\expandafter \@firstoftwo
 \else \expandafter \@secondoftwo
 \fi
}%
\providecommand \natexlab [1]{#1}%
\providecommand \enquote  [1]{``#1''}%
\providecommand \bibnamefont  [1]{#1}%
\providecommand \bibfnamefont [1]{#1}%
\providecommand \citenamefont [1]{#1}%
\providecommand \href@noop [0]{\@secondoftwo}%
\providecommand \href [0]{\begingroup \@sanitize@url \@href}%
\providecommand \@href[1]{\@@startlink{#1}\@@href}%
\providecommand \@@href[1]{\endgroup#1\@@endlink}%
\providecommand \@sanitize@url [0]{\catcode `\\12\catcode `\$12\catcode
  `\&12\catcode `\#12\catcode `\^12\catcode `\_12\catcode `\%12\relax}%
\providecommand \@@startlink[1]{}%
\providecommand \@@endlink[0]{}%
\providecommand \url  [0]{\begingroup\@sanitize@url \@url }%
\providecommand \@url [1]{\endgroup\@href {#1}{\urlprefix }}%
\providecommand \urlprefix  [0]{URL }%
\providecommand \Eprint [0]{\href }%
\providecommand \doibase [0]{https://doi.org/}%
\providecommand \selectlanguage [0]{\@gobble}%
\providecommand \bibinfo  [0]{\@secondoftwo}%
\providecommand \bibfield  [0]{\@secondoftwo}%
\providecommand \translation [1]{[#1]}%
\providecommand \BibitemOpen [0]{}%
\providecommand \bibitemStop [0]{}%
\providecommand \bibitemNoStop [0]{.\EOS\space}%
\providecommand \EOS [0]{\spacefactor3000\relax}%
\providecommand \BibitemShut  [1]{\csname bibitem#1\endcsname}%
\let\auto@bib@innerbib\@empty
%</preamble>
\bibitem [{\citenamefont {Martin}\ \emph {et~al.}(1990)\citenamefont {Martin},
  \citenamefont {Fiory}, \citenamefont {Fleming}, \citenamefont {Schneemeyer},\
  and\ \citenamefont {Waszczak}}]{martin1990normal}%
  \BibitemOpen
  \bibfield  {author} {\bibinfo {author} {\bibfnamefont {S.}~\bibnamefont
  {Martin}}, \bibinfo {author} {\bibfnamefont {A.~T.}\ \bibnamefont {Fiory}},
  \bibinfo {author} {\bibfnamefont {R.}~\bibnamefont {Fleming}}, \bibinfo
  {author} {\bibfnamefont {L.}~\bibnamefont {Schneemeyer}},\ and\ \bibinfo
  {author} {\bibfnamefont {J.~V.}\ \bibnamefont {Waszczak}},\ }\bibfield
  {title} {\bibinfo {title} {Normal-state transport properties of bi 2+ x sr 2-
  y cuo 6+ $\delta$ crystals},\ }\href@noop {} {\bibfield  {journal} {\bibinfo
  {journal} {Physical Review B}\ }\textbf {\bibinfo {volume} {41}},\ \bibinfo
  {pages} {846} (\bibinfo {year} {1990})}\BibitemShut {NoStop}%
\bibitem [{\citenamefont {Takagi}\ \emph {et~al.}(1992)\citenamefont {Takagi},
  \citenamefont {Batlogg}, \citenamefont {Kao}, \citenamefont {Kwo},
  \citenamefont {Cava}, \citenamefont {Krajewski},\ and\ \citenamefont
  {Peck~Jr}}]{takagi1992systematic}%
  \BibitemOpen
  \bibfield  {author} {\bibinfo {author} {\bibfnamefont {H.}~\bibnamefont
  {Takagi}}, \bibinfo {author} {\bibfnamefont {B.}~\bibnamefont {Batlogg}},
  \bibinfo {author} {\bibfnamefont {H.}~\bibnamefont {Kao}}, \bibinfo {author}
  {\bibfnamefont {J.}~\bibnamefont {Kwo}}, \bibinfo {author} {\bibfnamefont
  {R.~J.}\ \bibnamefont {Cava}}, \bibinfo {author} {\bibfnamefont
  {J.}~\bibnamefont {Krajewski}},\ and\ \bibinfo {author} {\bibfnamefont
  {W.}~\bibnamefont {Peck~Jr}},\ }\bibfield  {title} {\bibinfo {title}
  {Systematic evolution of temperature-dependent resistivity in la 2- x sr x
  cuo 4},\ }\href@noop {} {\bibfield  {journal} {\bibinfo  {journal} {Physical
  review letters}\ }\textbf {\bibinfo {volume} {69}},\ \bibinfo {pages} {2975}
  (\bibinfo {year} {1992})}\BibitemShut {NoStop}%
\bibitem [{\citenamefont {Zaanen}(2019)}]{zaanen2019planckian}%
  \BibitemOpen
  \bibfield  {author} {\bibinfo {author} {\bibfnamefont {J.}~\bibnamefont
  {Zaanen}},\ }\bibfield  {title} {\bibinfo {title} {Planckian dissipation,
  minimal viscosity and the transport in cuprate strange metals},\ }\href@noop
  {} {\bibfield  {journal} {\bibinfo  {journal} {SciPost Physics}\ }\textbf
  {\bibinfo {volume} {6}},\ \bibinfo {pages} {061} (\bibinfo {year}
  {2019})}\BibitemShut {NoStop}%
\bibitem [{\citenamefont {Hartnoll}(2015)}]{hartnoll2015theory}%
  \BibitemOpen
  \bibfield  {author} {\bibinfo {author} {\bibfnamefont {S.~A.}\ \bibnamefont
  {Hartnoll}},\ }\bibfield  {title} {\bibinfo {title} {Theory of universal
  incoherent metallic transport},\ }\href@noop {} {\bibfield  {journal}
  {\bibinfo  {journal} {Nature Physics}\ }\textbf {\bibinfo {volume} {11}},\
  \bibinfo {pages} {54} (\bibinfo {year} {2015})}\BibitemShut {NoStop}%
\bibitem [{\citenamefont {Bruin}\ \emph {et~al.}(2013)\citenamefont {Bruin},
  \citenamefont {Sakai}, \citenamefont {Perry},\ and\ \citenamefont
  {Mackenzie}}]{bruin2013similarity}%
  \BibitemOpen
  \bibfield  {author} {\bibinfo {author} {\bibfnamefont {J.}~\bibnamefont
  {Bruin}}, \bibinfo {author} {\bibfnamefont {H.}~\bibnamefont {Sakai}},
  \bibinfo {author} {\bibfnamefont {R.}~\bibnamefont {Perry}},\ and\ \bibinfo
  {author} {\bibfnamefont {A.}~\bibnamefont {Mackenzie}},\ }\bibfield  {title}
  {\bibinfo {title} {Similarity of scattering rates in metals showing t-linear
  resistivity},\ }\href@noop {} {\bibfield  {journal} {\bibinfo  {journal}
  {Science}\ }\textbf {\bibinfo {volume} {339}},\ \bibinfo {pages} {804}
  (\bibinfo {year} {2013})}\BibitemShut {NoStop}%
\bibitem [{\citenamefont {Legros}\ \emph {et~al.}(2019)\citenamefont {Legros},
  \citenamefont {Benhabib}, \citenamefont {Tabis}, \citenamefont
  {Lalibert{\'e}}, \citenamefont {Dion}, \citenamefont {Lizaire}, \citenamefont
  {Vignolle}, \citenamefont {Vignolles}, \citenamefont {Raffy}, \citenamefont
  {Li} \emph {et~al.}}]{legros2019universal}%
  \BibitemOpen
  \bibfield  {author} {\bibinfo {author} {\bibfnamefont {A.}~\bibnamefont
  {Legros}}, \bibinfo {author} {\bibfnamefont {S.}~\bibnamefont {Benhabib}},
  \bibinfo {author} {\bibfnamefont {W.}~\bibnamefont {Tabis}}, \bibinfo
  {author} {\bibfnamefont {F.}~\bibnamefont {Lalibert{\'e}}}, \bibinfo {author}
  {\bibfnamefont {M.}~\bibnamefont {Dion}}, \bibinfo {author} {\bibfnamefont
  {M.}~\bibnamefont {Lizaire}}, \bibinfo {author} {\bibfnamefont
  {B.}~\bibnamefont {Vignolle}}, \bibinfo {author} {\bibfnamefont
  {D.}~\bibnamefont {Vignolles}}, \bibinfo {author} {\bibfnamefont
  {H.}~\bibnamefont {Raffy}}, \bibinfo {author} {\bibfnamefont
  {Z.}~\bibnamefont {Li}}, \emph {et~al.},\ }\bibfield  {title} {\bibinfo
  {title} {Universal t-linear resistivity and planckian dissipation in
  overdoped cuprates},\ }\href@noop {} {\bibfield  {journal} {\bibinfo
  {journal} {Nature Physics}\ }\textbf {\bibinfo {volume} {15}},\ \bibinfo
  {pages} {142} (\bibinfo {year} {2019})}\BibitemShut {NoStop}%
\bibitem [{\citenamefont {Phillips}\ \emph {et~al.}(2022)\citenamefont
  {Phillips}, \citenamefont {Hussey},\ and\ \citenamefont
  {Abbamonte}}]{phillips2022stranger}%
  \BibitemOpen
  \bibfield  {author} {\bibinfo {author} {\bibfnamefont {P.~W.}\ \bibnamefont
  {Phillips}}, \bibinfo {author} {\bibfnamefont {N.~E.}\ \bibnamefont
  {Hussey}},\ and\ \bibinfo {author} {\bibfnamefont {P.}~\bibnamefont
  {Abbamonte}},\ }\bibfield  {title} {\bibinfo {title} {Stranger than metals},\
  }\href@noop {} {\bibfield  {journal} {\bibinfo  {journal} {Science}\ }\textbf
  {\bibinfo {volume} {377}},\ \bibinfo {pages} {eabh4273} (\bibinfo {year}
  {2022})}\BibitemShut {NoStop}%
\bibitem [{\citenamefont {Varma}\ \emph {et~al.}(1989)\citenamefont {Varma},
  \citenamefont {Littlewood}, \citenamefont {Schmitt-Rink}, \citenamefont
  {Abrahams},\ and\ \citenamefont {Ruckenstein}}]{varma1989phenomenology}%
  \BibitemOpen
  \bibfield  {author} {\bibinfo {author} {\bibfnamefont {C.}~\bibnamefont
  {Varma}}, \bibinfo {author} {\bibfnamefont {P.~B.}\ \bibnamefont
  {Littlewood}}, \bibinfo {author} {\bibfnamefont {S.}~\bibnamefont
  {Schmitt-Rink}}, \bibinfo {author} {\bibfnamefont {E.}~\bibnamefont
  {Abrahams}},\ and\ \bibinfo {author} {\bibfnamefont {A.}~\bibnamefont
  {Ruckenstein}},\ }\bibfield  {title} {\bibinfo {title} {Phenomenology of the
  normal state of cu-o high-temperature superconductors},\ }\href@noop {}
  {\bibfield  {journal} {\bibinfo  {journal} {Physical Review Letters}\
  }\textbf {\bibinfo {volume} {63}},\ \bibinfo {pages} {1996} (\bibinfo {year}
  {1989})}\BibitemShut {NoStop}%
\bibitem [{\citenamefont {Littlewood}\ and\ \citenamefont
  {Varma}(1991)}]{littlewood1991phenomenology}%
  \BibitemOpen
  \bibfield  {author} {\bibinfo {author} {\bibfnamefont {P.}~\bibnamefont
  {Littlewood}}\ and\ \bibinfo {author} {\bibfnamefont {C.}~\bibnamefont
  {Varma}},\ }\bibfield  {title} {\bibinfo {title} {Phenomenology of the normal
  and superconducting states of a marginal fermi liquid},\ }\href@noop {}
  {\bibfield  {journal} {\bibinfo  {journal} {Journal of applied physics}\
  }\textbf {\bibinfo {volume} {69}},\ \bibinfo {pages} {4979} (\bibinfo {year}
  {1991})}\BibitemShut {NoStop}%
\bibitem [{\citenamefont {Vishik}\ \emph {et~al.}(2010)\citenamefont {Vishik},
  \citenamefont {Lee}, \citenamefont {He}, \citenamefont {Hashimoto},
  \citenamefont {Hussain}, \citenamefont {Devereaux},\ and\ \citenamefont
  {Shen}}]{Vishik2010}%
  \BibitemOpen
  \bibfield  {author} {\bibinfo {author} {\bibfnamefont {I.~M.}\ \bibnamefont
  {Vishik}}, \bibinfo {author} {\bibfnamefont {W.~S.}\ \bibnamefont {Lee}},
  \bibinfo {author} {\bibfnamefont {R.-H.}\ \bibnamefont {He}}, \bibinfo
  {author} {\bibfnamefont {M.}~\bibnamefont {Hashimoto}}, \bibinfo {author}
  {\bibfnamefont {Z.}~\bibnamefont {Hussain}}, \bibinfo {author} {\bibfnamefont
  {T.~P.}\ \bibnamefont {Devereaux}},\ and\ \bibinfo {author} {\bibfnamefont
  {Z.-X.}\ \bibnamefont {Shen}},\ }\bibfield  {title} {\bibinfo {title} {Arpes
  studies of cuprate fermiology: superconductivity, pseudogap and quasiparticle
  dynamics},\ }\href {https://doi.org/10.1088/1367-2630/12/10/105008}
  {\bibfield  {journal} {\bibinfo  {journal} {New Journal of Physics}\ }\textbf
  {\bibinfo {volume} {12}},\ \bibinfo {pages} {105008} (\bibinfo {year}
  {2010})}\BibitemShut {NoStop}%
\bibitem [{\citenamefont {Reber}\ \emph {et~al.}(2019)\citenamefont {Reber},
  \citenamefont {Zhou}, \citenamefont {Plumb}, \citenamefont {Parham},
  \citenamefont {Waugh}, \citenamefont {Cao}, \citenamefont {Sun},
  \citenamefont {Li}, \citenamefont {Wang}, \citenamefont {Wen} \emph
  {et~al.}}]{reber2019unified}%
  \BibitemOpen
  \bibfield  {author} {\bibinfo {author} {\bibfnamefont {T.~J.}\ \bibnamefont
  {Reber}}, \bibinfo {author} {\bibfnamefont {X.}~\bibnamefont {Zhou}},
  \bibinfo {author} {\bibfnamefont {N.}~\bibnamefont {Plumb}}, \bibinfo
  {author} {\bibfnamefont {S.}~\bibnamefont {Parham}}, \bibinfo {author}
  {\bibfnamefont {J.}~\bibnamefont {Waugh}}, \bibinfo {author} {\bibfnamefont
  {Y.}~\bibnamefont {Cao}}, \bibinfo {author} {\bibfnamefont {Z.}~\bibnamefont
  {Sun}}, \bibinfo {author} {\bibfnamefont {H.}~\bibnamefont {Li}}, \bibinfo
  {author} {\bibfnamefont {Q.}~\bibnamefont {Wang}}, \bibinfo {author}
  {\bibfnamefont {J.}~\bibnamefont {Wen}}, \emph {et~al.},\ }\bibfield  {title}
  {\bibinfo {title} {A unified form of low-energy nodal electronic interactions
  in hole-doped cuprate superconductors},\ }\href@noop {} {\bibfield  {journal}
  {\bibinfo  {journal} {Nature Communications}\ }\textbf {\bibinfo {volume}
  {10}},\ \bibinfo {pages} {5737} (\bibinfo {year} {2019})}\BibitemShut
  {NoStop}%
\bibitem [{\citenamefont {Levallois}\ \emph {et~al.}(2016)\citenamefont
  {Levallois}, \citenamefont {Tran}, \citenamefont {Pouliot}, \citenamefont
  {Presura}, \citenamefont {Greene}, \citenamefont {Eckstein}, \citenamefont
  {Uccelli}, \citenamefont {Giannini}, \citenamefont {Gu}, \citenamefont
  {Leggett} \emph {et~al.}}]{levallois2016temperature}%
  \BibitemOpen
  \bibfield  {author} {\bibinfo {author} {\bibfnamefont {J.}~\bibnamefont
  {Levallois}}, \bibinfo {author} {\bibfnamefont {M.}~\bibnamefont {Tran}},
  \bibinfo {author} {\bibfnamefont {D.}~\bibnamefont {Pouliot}}, \bibinfo
  {author} {\bibfnamefont {C.}~\bibnamefont {Presura}}, \bibinfo {author}
  {\bibfnamefont {L.}~\bibnamefont {Greene}}, \bibinfo {author} {\bibfnamefont
  {J.}~\bibnamefont {Eckstein}}, \bibinfo {author} {\bibfnamefont
  {J.}~\bibnamefont {Uccelli}}, \bibinfo {author} {\bibfnamefont
  {E.}~\bibnamefont {Giannini}}, \bibinfo {author} {\bibfnamefont
  {G.}~\bibnamefont {Gu}}, \bibinfo {author} {\bibfnamefont {A.}~\bibnamefont
  {Leggett}}, \emph {et~al.},\ }\bibfield  {title} {\bibinfo {title}
  {Temperature-dependent ellipsometry measurements of partial coulomb energy in
  superconducting cuprates},\ }\href@noop {} {\bibfield  {journal} {\bibinfo
  {journal} {Physical Review X}\ }\textbf {\bibinfo {volume} {6}},\ \bibinfo
  {pages} {031027} (\bibinfo {year} {2016})}\BibitemShut {NoStop}%
\bibitem [{\citenamefont {Marel}\ \emph {et~al.}(2003)\citenamefont {Marel},
  \citenamefont {Molegraaf}, \citenamefont {Zaanen}, \citenamefont {Nussinov},
  \citenamefont {Carbone}, \citenamefont {Damascelli}, \citenamefont {Eisaki},
  \citenamefont {Greven}, \citenamefont {Kes},\ and\ \citenamefont
  {Li}}]{marel2003quantum}%
  \BibitemOpen
  \bibfield  {author} {\bibinfo {author} {\bibfnamefont {D.~v.~d.}\
  \bibnamefont {Marel}}, \bibinfo {author} {\bibfnamefont {H.}~\bibnamefont
  {Molegraaf}}, \bibinfo {author} {\bibfnamefont {J.}~\bibnamefont {Zaanen}},
  \bibinfo {author} {\bibfnamefont {Z.}~\bibnamefont {Nussinov}}, \bibinfo
  {author} {\bibfnamefont {F.}~\bibnamefont {Carbone}}, \bibinfo {author}
  {\bibfnamefont {A.}~\bibnamefont {Damascelli}}, \bibinfo {author}
  {\bibfnamefont {H.}~\bibnamefont {Eisaki}}, \bibinfo {author} {\bibfnamefont
  {M.}~\bibnamefont {Greven}}, \bibinfo {author} {\bibfnamefont
  {P.}~\bibnamefont {Kes}},\ and\ \bibinfo {author} {\bibfnamefont
  {M.}~\bibnamefont {Li}},\ }\bibfield  {title} {\bibinfo {title} {Quantum
  critical behaviour in a high-t c superconductor},\ }\href@noop {} {\bibfield
  {journal} {\bibinfo  {journal} {Nature}\ }\textbf {\bibinfo {volume} {425}},\
  \bibinfo {pages} {271} (\bibinfo {year} {2003})}\BibitemShut {NoStop}%
\bibitem [{\citenamefont {Mitrano}\ \emph {et~al.}(2018)\citenamefont
  {Mitrano}, \citenamefont {Husain}, \citenamefont {Vig}, \citenamefont
  {Kogar}, \citenamefont {Rak}, \citenamefont {Rubeck}, \citenamefont
  {Schmalian}, \citenamefont {Uchoa}, \citenamefont {Schneeloch}, \citenamefont
  {Zhong} \emph {et~al.}}]{mitrano2018anomalous}%
  \BibitemOpen
  \bibfield  {author} {\bibinfo {author} {\bibfnamefont {M.}~\bibnamefont
  {Mitrano}}, \bibinfo {author} {\bibfnamefont {A.}~\bibnamefont {Husain}},
  \bibinfo {author} {\bibfnamefont {S.}~\bibnamefont {Vig}}, \bibinfo {author}
  {\bibfnamefont {A.}~\bibnamefont {Kogar}}, \bibinfo {author} {\bibfnamefont
  {M.}~\bibnamefont {Rak}}, \bibinfo {author} {\bibfnamefont {S.}~\bibnamefont
  {Rubeck}}, \bibinfo {author} {\bibfnamefont {J.}~\bibnamefont {Schmalian}},
  \bibinfo {author} {\bibfnamefont {B.}~\bibnamefont {Uchoa}}, \bibinfo
  {author} {\bibfnamefont {J.}~\bibnamefont {Schneeloch}}, \bibinfo {author}
  {\bibfnamefont {R.}~\bibnamefont {Zhong}}, \emph {et~al.},\ }\bibfield
  {title} {\bibinfo {title} {Anomalous density fluctuations in a strange
  metal},\ }\href@noop {} {\bibfield  {journal} {\bibinfo  {journal}
  {Proceedings of the National Academy of Sciences}\ }\textbf {\bibinfo
  {volume} {115}},\ \bibinfo {pages} {5392} (\bibinfo {year}
  {2018})}\BibitemShut {NoStop}%
\bibitem [{\citenamefont {Husain}\ \emph {et~al.}(2019)\citenamefont {Husain},
  \citenamefont {Mitrano}, \citenamefont {Rak}, \citenamefont {Rubeck},
  \citenamefont {Uchoa}, \citenamefont {March}, \citenamefont {Dwyer},
  \citenamefont {Schneeloch}, \citenamefont {Zhong}, \citenamefont {Gu} \emph
  {et~al.}}]{husain2019crossover}%
  \BibitemOpen
  \bibfield  {author} {\bibinfo {author} {\bibfnamefont {A.~A.}\ \bibnamefont
  {Husain}}, \bibinfo {author} {\bibfnamefont {M.}~\bibnamefont {Mitrano}},
  \bibinfo {author} {\bibfnamefont {M.~S.}\ \bibnamefont {Rak}}, \bibinfo
  {author} {\bibfnamefont {S.}~\bibnamefont {Rubeck}}, \bibinfo {author}
  {\bibfnamefont {B.}~\bibnamefont {Uchoa}}, \bibinfo {author} {\bibfnamefont
  {K.}~\bibnamefont {March}}, \bibinfo {author} {\bibfnamefont
  {C.}~\bibnamefont {Dwyer}}, \bibinfo {author} {\bibfnamefont
  {J.}~\bibnamefont {Schneeloch}}, \bibinfo {author} {\bibfnamefont
  {R.}~\bibnamefont {Zhong}}, \bibinfo {author} {\bibfnamefont {G.~D.}\
  \bibnamefont {Gu}}, \emph {et~al.},\ }\bibfield  {title} {\bibinfo {title}
  {Crossover of charge fluctuations across the strange metal phase diagram},\
  }\href@noop {} {\bibfield  {journal} {\bibinfo  {journal} {Physical Review
  X}\ }\textbf {\bibinfo {volume} {9}},\ \bibinfo {pages} {041062} (\bibinfo
  {year} {2019})}\BibitemShut {NoStop}%
\bibitem [{\citenamefont {Schulte}(2002)}]{schulte2002interplay}%
  \BibitemOpen
  \bibfield  {author} {\bibinfo {author} {\bibfnamefont {K.~H.~G.}\
  \bibnamefont {Schulte}},\ }\emph {\bibinfo {title} {The interplay of
  spectroscopy and correlated materials}},\ \href@noop {} {Ph.D. thesis},\
  \bibinfo  {school} {University of Groningen} (\bibinfo {year}
  {2002})\BibitemShut {NoStop}%
\bibitem [{\citenamefont {Basov}\ and\ \citenamefont
  {Timusk}(2005)}]{basov2005electrodynamics}%
  \BibitemOpen
  \bibfield  {author} {\bibinfo {author} {\bibfnamefont {D.}~\bibnamefont
  {Basov}}\ and\ \bibinfo {author} {\bibfnamefont {T.}~\bibnamefont {Timusk}},\
  }\bibfield  {title} {\bibinfo {title} {Electrodynamics of high-t c
  superconductors},\ }\href@noop {} {\bibfield  {journal} {\bibinfo  {journal}
  {Reviews of modern physics}\ }\textbf {\bibinfo {volume} {77}},\ \bibinfo
  {pages} {721} (\bibinfo {year} {2005})}\BibitemShut {NoStop}%
\bibitem [{\citenamefont {Evans}\ and\ \citenamefont
  {Mills}(1972)}]{evans1972theory}%
  \BibitemOpen
  \bibfield  {author} {\bibinfo {author} {\bibfnamefont {E.}~\bibnamefont
  {Evans}}\ and\ \bibinfo {author} {\bibfnamefont {D.}~\bibnamefont {Mills}},\
  }\bibfield  {title} {\bibinfo {title} {Theory of inelastic scattering of slow
  electrons by long-wavelength surface optical phonons},\ }\href@noop {}
  {\bibfield  {journal} {\bibinfo  {journal} {Physical Review B}\ }\textbf
  {\bibinfo {volume} {5}},\ \bibinfo {pages} {4126} (\bibinfo {year}
  {1972})}\BibitemShut {NoStop}%
\bibitem [{\citenamefont {Mills}(1975)}]{mills1975scattering}%
  \BibitemOpen
  \bibfield  {author} {\bibinfo {author} {\bibfnamefont {D.}~\bibnamefont
  {Mills}},\ }\bibfield  {title} {\bibinfo {title} {The scattering of low
  energy electrons by electric field fluctuations near crystal surfaces},\
  }\href@noop {} {\bibfield  {journal} {\bibinfo  {journal} {Surface Science}\
  }\textbf {\bibinfo {volume} {48}},\ \bibinfo {pages} {59} (\bibinfo {year}
  {1975})}\BibitemShut {NoStop}%
\bibitem [{\citenamefont {Ibach}\ and\ \citenamefont
  {Mills}(2013)}]{ibach2013electron}%
  \BibitemOpen
  \bibfield  {author} {\bibinfo {author} {\bibfnamefont {H.}~\bibnamefont
  {Ibach}}\ and\ \bibinfo {author} {\bibfnamefont {D.~L.}\ \bibnamefont
  {Mills}},\ }\href@noop {} {\emph {\bibinfo {title} {Electron energy loss
  spectroscopy and surface vibrations}}}\ (\bibinfo  {publisher} {Academic
  press},\ \bibinfo {year} {2013})\BibitemShut {NoStop}%
\bibitem [{\citenamefont {Vig}\ \emph {et~al.}(2017)\citenamefont {Vig},
  \citenamefont {Kogar}, \citenamefont {Mitrano}, \citenamefont {Husain},
  \citenamefont {Venema}, \citenamefont {Rak}, \citenamefont {Mishra},
  \citenamefont {Johnson}, \citenamefont {Gu}, \citenamefont {Fradkin} \emph
  {et~al.}}]{vig2017measurement}%
  \BibitemOpen
  \bibfield  {author} {\bibinfo {author} {\bibfnamefont {S.}~\bibnamefont
  {Vig}}, \bibinfo {author} {\bibfnamefont {A.}~\bibnamefont {Kogar}}, \bibinfo
  {author} {\bibfnamefont {M.}~\bibnamefont {Mitrano}}, \bibinfo {author}
  {\bibfnamefont {A.}~\bibnamefont {Husain}}, \bibinfo {author} {\bibfnamefont
  {L.}~\bibnamefont {Venema}}, \bibinfo {author} {\bibfnamefont
  {M.}~\bibnamefont {Rak}}, \bibinfo {author} {\bibfnamefont {V.}~\bibnamefont
  {Mishra}}, \bibinfo {author} {\bibfnamefont {P.}~\bibnamefont {Johnson}},
  \bibinfo {author} {\bibfnamefont {G.}~\bibnamefont {Gu}}, \bibinfo {author}
  {\bibfnamefont {E.}~\bibnamefont {Fradkin}}, \emph {et~al.},\ }\bibfield
  {title} {\bibinfo {title} {Measurement of the dynamic charge response of
  materials using low-energy, momentum-resolved electron energy-loss
  spectroscopy (m-eels)},\ }\href@noop {} {\bibfield  {journal} {\bibinfo
  {journal} {SciPost Physics}\ }\textbf {\bibinfo {volume} {3}},\ \bibinfo
  {pages} {026} (\bibinfo {year} {2017})}\BibitemShut {NoStop}%
\bibitem [{\citenamefont {Wen}\ \emph {et~al.}(2008)\citenamefont {Wen},
  \citenamefont {Xu}, \citenamefont {Xu}, \citenamefont {Hücker},
  \citenamefont {Tranquada},\ and\ \citenamefont {Gu}}]{wen2008}%
  \BibitemOpen
  \bibfield  {author} {\bibinfo {author} {\bibfnamefont {J.}~\bibnamefont
  {Wen}}, \bibinfo {author} {\bibfnamefont {Z.}~\bibnamefont {Xu}}, \bibinfo
  {author} {\bibfnamefont {G.}~\bibnamefont {Xu}}, \bibinfo {author}
  {\bibfnamefont {M.}~\bibnamefont {Hücker}}, \bibinfo {author} {\bibfnamefont
  {J.}~\bibnamefont {Tranquada}},\ and\ \bibinfo {author} {\bibfnamefont
  {G.}~\bibnamefont {Gu}},\ }\bibfield  {title} {\bibinfo {title} {Large
  bi-2212 single crystal growth by the floating-zone technique},\ }\href@noop
  {} {\bibfield  {journal} {\bibinfo  {journal} {Journal of Crystal Growth}\
  }\textbf {\bibinfo {volume} {310}},\ \bibinfo {pages} {1401} (\bibinfo {year}
  {2008})}\BibitemShut {NoStop}%
\bibitem [{\citenamefont {Presland}\ \emph {et~al.}(1991)\citenamefont
  {Presland}, \citenamefont {Tallon}, \citenamefont {Buckley}, \citenamefont
  {Liu},\ and\ \citenamefont {Flower}}]{presland1991general}%
  \BibitemOpen
  \bibfield  {author} {\bibinfo {author} {\bibfnamefont {M.}~\bibnamefont
  {Presland}}, \bibinfo {author} {\bibfnamefont {J.}~\bibnamefont {Tallon}},
  \bibinfo {author} {\bibfnamefont {R.}~\bibnamefont {Buckley}}, \bibinfo
  {author} {\bibfnamefont {R.}~\bibnamefont {Liu}},\ and\ \bibinfo {author}
  {\bibfnamefont {N.}~\bibnamefont {Flower}},\ }\bibfield  {title} {\bibinfo
  {title} {General trends in oxygen stoichiometry effects on tc in bi and tl
  superconductors},\ }\href@noop {} {\bibfield  {journal} {\bibinfo  {journal}
  {Physica C: Superconductivity}\ }\textbf {\bibinfo {volume} {176}},\ \bibinfo
  {pages} {95} (\bibinfo {year} {1991})}\BibitemShut {NoStop}%
\bibitem [{\citenamefont {Damascelli}\ \emph {et~al.}(2003)\citenamefont
  {Damascelli}, \citenamefont {Hussain},\ and\ \citenamefont
  {Shen}}]{damascelli2003}%
  \BibitemOpen
  \bibfield  {author} {\bibinfo {author} {\bibfnamefont {A.}~\bibnamefont
  {Damascelli}}, \bibinfo {author} {\bibfnamefont {Z.}~\bibnamefont
  {Hussain}},\ and\ \bibinfo {author} {\bibfnamefont {Z.-X.}\ \bibnamefont
  {Shen}},\ }\bibfield  {title} {\bibinfo {title} {Angle-resolved photoemission
  studies of the cuprate superconductors},\ }\href@noop {} {\bibfield
  {journal} {\bibinfo  {journal} {Rev. Mod. Phys.}\ }\textbf {\bibinfo {volume}
  {75}},\ \bibinfo {pages} {473} (\bibinfo {year} {2003})}\BibitemShut
  {NoStop}%
\bibitem [{Ere()}]{EresNote}%
  \BibitemOpen
  \href@noop {} {}\bibinfo {note} {We neglect the small contribution from the
  finite energy resolution of the instrument, since $\Delta E/E \sim
  10^{-4}$.}\BibitemShut {Stop}%
\bibitem [{\citenamefont {Li}\ \emph {et~al.}(2022)\citenamefont {Li},
  \citenamefont {Lin}, \citenamefont {Miao}, \citenamefont {Zhong},
  \citenamefont {Xue}, \citenamefont {Li}, \citenamefont {Tao}, \citenamefont
  {Wang}, \citenamefont {Guo},\ and\ \citenamefont {Zhu}}]{li2022geometric}%
  \BibitemOpen
  \bibfield  {author} {\bibinfo {author} {\bibfnamefont {J.}~\bibnamefont
  {Li}}, \bibinfo {author} {\bibfnamefont {Z.}~\bibnamefont {Lin}}, \bibinfo
  {author} {\bibfnamefont {G.}~\bibnamefont {Miao}}, \bibinfo {author}
  {\bibfnamefont {W.}~\bibnamefont {Zhong}}, \bibinfo {author} {\bibfnamefont
  {S.}~\bibnamefont {Xue}}, \bibinfo {author} {\bibfnamefont {Y.}~\bibnamefont
  {Li}}, \bibinfo {author} {\bibfnamefont {Z.}~\bibnamefont {Tao}}, \bibinfo
  {author} {\bibfnamefont {W.}~\bibnamefont {Wang}}, \bibinfo {author}
  {\bibfnamefont {J.}~\bibnamefont {Guo}},\ and\ \bibinfo {author}
  {\bibfnamefont {X.}~\bibnamefont {Zhu}},\ }\bibfield  {title} {\bibinfo
  {title} {Geometric effect of high-resolution electron energy loss
  spectroscopy on the identification of plasmons: An example of graphene},\
  }\href@noop {} {\bibfield  {journal} {\bibinfo  {journal} {Surface Science}\
  }\textbf {\bibinfo {volume} {721}},\ \bibinfo {pages} {122067} (\bibinfo
  {year} {2022})}\BibitemShut {NoStop}%
\bibitem [{\citenamefont {Jain}\ and\ \citenamefont
  {Allen}(1985)}]{jain1985dielectric}%
  \BibitemOpen
  \bibfield  {author} {\bibinfo {author} {\bibfnamefont {J.~K.}\ \bibnamefont
  {Jain}}\ and\ \bibinfo {author} {\bibfnamefont {P.~B.}\ \bibnamefont
  {Allen}},\ }\bibfield  {title} {\bibinfo {title} {Dielectric response of a
  semi-infinite layered electron gas and raman scattering from its bulk and
  surface plasmons},\ }\href@noop {} {\bibfield  {journal} {\bibinfo  {journal}
  {Physical Review B}\ }\textbf {\bibinfo {volume} {32}},\ \bibinfo {pages}
  {997} (\bibinfo {year} {1985})}\BibitemShut {NoStop}%
\bibitem [{\citenamefont {Olego}\ \emph {et~al.}(1982)\citenamefont {Olego},
  \citenamefont {Pinczuk}, \citenamefont {Gossard},\ and\ \citenamefont
  {Wiegmann}}]{Olego1982}%
  \BibitemOpen
  \bibfield  {author} {\bibinfo {author} {\bibfnamefont {D.}~\bibnamefont
  {Olego}}, \bibinfo {author} {\bibfnamefont {A.}~\bibnamefont {Pinczuk}},
  \bibinfo {author} {\bibfnamefont {A.~C.}\ \bibnamefont {Gossard}},\ and\
  \bibinfo {author} {\bibfnamefont {W.}~\bibnamefont {Wiegmann}},\ }\bibfield
  {title} {\bibinfo {title} {Plasma dispersion in a layered electron gas: A
  determination in gaas-(alga) as heterostructures},\ }\href
  {https://doi.org/10.1103/PhysRevB.25.7867} {\bibfield  {journal} {\bibinfo
  {journal} {Phys. Rev. B}\ }\textbf {\bibinfo {volume} {25}},\ \bibinfo
  {pages} {7867} (\bibinfo {year} {1982})}\BibitemShut {NoStop}%
\bibitem [{bil()}]{bilayerNote}%
  \BibitemOpen
  \href@noop {} {}\bibinfo {note} {We treat the CuO$_2$ bilayer here as a
  single structural unit, which is valid since the bilayer splitting in Bi-2212
  is much smaller than the plasma frequency.}\BibitemShut {Stop}%
\bibitem [{\citenamefont {Pines}\ and\ \citenamefont
  {Nozi\`eres}(1999)}]{PinesNozieres1973}%
  \BibitemOpen
  \bibfield  {author} {\bibinfo {author} {\bibfnamefont {D.}~\bibnamefont
  {Pines}}\ and\ \bibinfo {author} {\bibfnamefont {P.}~\bibnamefont
  {Nozi\`eres}},\ }\href@noop {} {\emph {\bibinfo {title} {The Theory of
  Quantum Liquids}}}\ (\bibinfo  {publisher} {Perseus Books, Cambridge, MA},\
  \bibinfo {year} {1999})\BibitemShut {NoStop}%
\bibitem [{\citenamefont {Fetter}(1974)}]{fetter1974electrodynamics}%
  \BibitemOpen
  \bibfield  {author} {\bibinfo {author} {\bibfnamefont {A.~L.}\ \bibnamefont
  {Fetter}},\ }\bibfield  {title} {\bibinfo {title} {Electrodynamics of a
  layered electron gas. ii. periodic array},\ }\href@noop {} {\bibfield
  {journal} {\bibinfo  {journal} {Annals of Physics}\ }\textbf {\bibinfo
  {volume} {88}},\ \bibinfo {pages} {1} (\bibinfo {year} {1974})}\BibitemShut
  {NoStop}%
\bibitem [{\citenamefont {Bozovic}(1990)}]{bozovic1990plasmons}%
  \BibitemOpen
  \bibfield  {author} {\bibinfo {author} {\bibfnamefont {I.}~\bibnamefont
  {Bozovic}},\ }\bibfield  {title} {\bibinfo {title} {Plasmons in cuprate
  superconductors},\ }\href@noop {} {\bibfield  {journal} {\bibinfo  {journal}
  {Physical Review B}\ }\textbf {\bibinfo {volume} {42}},\ \bibinfo {pages}
  {1969} (\bibinfo {year} {1990})}\BibitemShut {NoStop}%
\bibitem [{\citenamefont {Nag}\ \emph {et~al.}(2020)\citenamefont {Nag},
  \citenamefont {Zhu}, \citenamefont {Bejas}, \citenamefont {Li}, \citenamefont
  {Robarts}, \citenamefont {Yamase}, \citenamefont {Petsch}, \citenamefont
  {Song}, \citenamefont {Eisaki}, \citenamefont {Walters} \emph
  {et~al.}}]{nag2020detection}%
  \BibitemOpen
  \bibfield  {author} {\bibinfo {author} {\bibfnamefont {A.}~\bibnamefont
  {Nag}}, \bibinfo {author} {\bibfnamefont {M.}~\bibnamefont {Zhu}}, \bibinfo
  {author} {\bibfnamefont {M.}~\bibnamefont {Bejas}}, \bibinfo {author}
  {\bibfnamefont {J.}~\bibnamefont {Li}}, \bibinfo {author} {\bibfnamefont
  {H.}~\bibnamefont {Robarts}}, \bibinfo {author} {\bibfnamefont
  {H.}~\bibnamefont {Yamase}}, \bibinfo {author} {\bibfnamefont
  {A.}~\bibnamefont {Petsch}}, \bibinfo {author} {\bibfnamefont
  {D.}~\bibnamefont {Song}}, \bibinfo {author} {\bibfnamefont {H.}~\bibnamefont
  {Eisaki}}, \bibinfo {author} {\bibfnamefont {A.}~\bibnamefont {Walters}},
  \emph {et~al.},\ }\bibfield  {title} {\bibinfo {title} {Detection of acoustic
  plasmons in hole-doped lanthanum and bismuth cuprate superconductors using
  resonant inelastic x-ray scattering},\ }\href@noop {} {\bibfield  {journal}
  {\bibinfo  {journal} {Physical Review Letters}\ }\textbf {\bibinfo {volume}
  {125}},\ \bibinfo {pages} {257002} (\bibinfo {year} {2020})}\BibitemShut
  {NoStop}%
\bibitem [{\citenamefont {Hepting}\ \emph {et~al.}(2018)\citenamefont
  {Hepting}, \citenamefont {Chaix}, \citenamefont {Huang}, \citenamefont
  {Fumagalli}, \citenamefont {Peng}, \citenamefont {Moritz}, \citenamefont
  {Kummer}, \citenamefont {Brookes}, \citenamefont {Lee}, \citenamefont
  {Hashimoto} \emph {et~al.}}]{hepting2018three}%
  \BibitemOpen
  \bibfield  {author} {\bibinfo {author} {\bibfnamefont {M.}~\bibnamefont
  {Hepting}}, \bibinfo {author} {\bibfnamefont {L.}~\bibnamefont {Chaix}},
  \bibinfo {author} {\bibfnamefont {E.}~\bibnamefont {Huang}}, \bibinfo
  {author} {\bibfnamefont {R.}~\bibnamefont {Fumagalli}}, \bibinfo {author}
  {\bibfnamefont {Y.}~\bibnamefont {Peng}}, \bibinfo {author} {\bibfnamefont
  {B.}~\bibnamefont {Moritz}}, \bibinfo {author} {\bibfnamefont
  {K.}~\bibnamefont {Kummer}}, \bibinfo {author} {\bibfnamefont
  {N.}~\bibnamefont {Brookes}}, \bibinfo {author} {\bibfnamefont
  {W.}~\bibnamefont {Lee}}, \bibinfo {author} {\bibfnamefont {M.}~\bibnamefont
  {Hashimoto}}, \emph {et~al.},\ }\bibfield  {title} {\bibinfo {title}
  {Three-dimensional collective charge excitations in electron-doped copper
  oxide superconductors},\ }\href@noop {} {\bibfield  {journal} {\bibinfo
  {journal} {Nature}\ }\textbf {\bibinfo {volume} {563}},\ \bibinfo {pages}
  {374} (\bibinfo {year} {2018})}\BibitemShut {NoStop}%
\bibitem [{\citenamefont {Hepting}\ \emph {et~al.}(2022)\citenamefont
  {Hepting}, \citenamefont {Bejas}, \citenamefont {Nag}, \citenamefont
  {Yamase}, \citenamefont {Coppola}, \citenamefont {Betto}, \citenamefont
  {Falter}, \citenamefont {Garcia-Fernandez}, \citenamefont {Agrestini},
  \citenamefont {Zhou} \emph {et~al.}}]{hepting2022gapped}%
  \BibitemOpen
  \bibfield  {author} {\bibinfo {author} {\bibfnamefont {M.}~\bibnamefont
  {Hepting}}, \bibinfo {author} {\bibfnamefont {M.}~\bibnamefont {Bejas}},
  \bibinfo {author} {\bibfnamefont {A.}~\bibnamefont {Nag}}, \bibinfo {author}
  {\bibfnamefont {H.}~\bibnamefont {Yamase}}, \bibinfo {author} {\bibfnamefont
  {N.}~\bibnamefont {Coppola}}, \bibinfo {author} {\bibfnamefont
  {D.}~\bibnamefont {Betto}}, \bibinfo {author} {\bibfnamefont
  {C.}~\bibnamefont {Falter}}, \bibinfo {author} {\bibfnamefont
  {M.}~\bibnamefont {Garcia-Fernandez}}, \bibinfo {author} {\bibfnamefont
  {S.}~\bibnamefont {Agrestini}}, \bibinfo {author} {\bibfnamefont {K.-J.}\
  \bibnamefont {Zhou}}, \emph {et~al.},\ }\bibfield  {title} {\bibinfo {title}
  {Gapped collective charge excitations and interlayer hopping in cuprate
  superconductors},\ }\href@noop {} {\bibfield  {journal} {\bibinfo  {journal}
  {Physical Review Letters}\ }\textbf {\bibinfo {volume} {129}},\ \bibinfo
  {pages} {047001} (\bibinfo {year} {2022})}\BibitemShut {NoStop}%
\bibitem [{\citenamefont {Warren}\ \emph {et~al.}(1980)\citenamefont {Warren},
  \citenamefont {Weitekamp},\ and\ \citenamefont {Pines}}]{warren1980theory}%
  \BibitemOpen
  \bibfield  {author} {\bibinfo {author} {\bibfnamefont {W.}~\bibnamefont
  {Warren}}, \bibinfo {author} {\bibfnamefont {D.}~\bibnamefont {Weitekamp}},\
  and\ \bibinfo {author} {\bibfnamefont {A.}~\bibnamefont {Pines}},\ }\bibfield
   {title} {\bibinfo {title} {Theory of selective excitation of
  multiple-quantum transitions},\ }\href@noop {} {\bibfield  {journal}
  {\bibinfo  {journal} {The Journal of Chemical Physics}\ }\textbf {\bibinfo
  {volume} {73}},\ \bibinfo {pages} {2084} (\bibinfo {year}
  {1980})}\BibitemShut {NoStop}%
\bibitem [{\citenamefont {Garciia~de Abajo}(2013)}]{garciia2013multiple}%
  \BibitemOpen
  \bibfield  {author} {\bibinfo {author} {\bibfnamefont {F.~J.}\ \bibnamefont
  {Garciia~de Abajo}},\ }\bibfield  {title} {\bibinfo {title} {Multiple
  excitation of confined graphene plasmons by single free electrons},\
  }\href@noop {} {\bibfield  {journal} {\bibinfo  {journal} {ACS nano}\
  }\textbf {\bibinfo {volume} {7}},\ \bibinfo {pages} {11409} (\bibinfo {year}
  {2013})}\BibitemShut {NoStop}%
\bibitem [{\citenamefont {N\"ucker}\ \emph {et~al.}(1989)\citenamefont
  {N\"ucker}, \citenamefont {Romberg}, \citenamefont {Nakai}, \citenamefont
  {Scheerer}, \citenamefont {Fink}, \citenamefont {Yan},\ and\ \citenamefont
  {Zhao}}]{Nucker1989}%
  \BibitemOpen
  \bibfield  {author} {\bibinfo {author} {\bibfnamefont {N.}~\bibnamefont
  {N\"ucker}}, \bibinfo {author} {\bibfnamefont {H.}~\bibnamefont {Romberg}},
  \bibinfo {author} {\bibfnamefont {S.}~\bibnamefont {Nakai}}, \bibinfo
  {author} {\bibfnamefont {B.}~\bibnamefont {Scheerer}}, \bibinfo {author}
  {\bibfnamefont {J.}~\bibnamefont {Fink}}, \bibinfo {author} {\bibfnamefont
  {Y.~F.}\ \bibnamefont {Yan}},\ and\ \bibinfo {author} {\bibfnamefont {Z.~X.}\
  \bibnamefont {Zhao}},\ }\bibfield  {title} {\bibinfo {title} {Plasmons and
  interband transitions in
  ${\mathrm{bi}}_{2}$${\mathrm{sr}}_{2}$ca${\mathrm{cu}}_{2}$${\mathrm{o}}_{8}$},\
  }\href {https://doi.org/10.1103/PhysRevB.39.12379} {\bibfield  {journal}
  {\bibinfo  {journal} {Phys. Rev. B}\ }\textbf {\bibinfo {volume} {39}},\
  \bibinfo {pages} {12379} (\bibinfo {year} {1989})}\BibitemShut {NoStop}%
\bibitem [{\citenamefont {Wang}\ \emph {et~al.}(1990)\citenamefont {Wang},
  \citenamefont {Feng},\ and\ \citenamefont {Ritter}}]{wang1990electron}%
  \BibitemOpen
  \bibfield  {author} {\bibinfo {author} {\bibfnamefont {Y.-Y.}\ \bibnamefont
  {Wang}}, \bibinfo {author} {\bibfnamefont {G.}~\bibnamefont {Feng}},\ and\
  \bibinfo {author} {\bibfnamefont {A.~L.}\ \bibnamefont {Ritter}},\ }\bibfield
   {title} {\bibinfo {title} {Electron-energy-loss and optical-transmittance
  investigation of bi 2 sr 2 cacu 2 o 8},\ }\href@noop {} {\bibfield  {journal}
  {\bibinfo  {journal} {Physical Review B}\ }\textbf {\bibinfo {volume} {42}},\
  \bibinfo {pages} {420} (\bibinfo {year} {1990})}\BibitemShut {NoStop}%
\bibitem [{\citenamefont {Terauchi}\ \emph {et~al.}(1995)\citenamefont
  {Terauchi}, \citenamefont {Tanaka}, \citenamefont {Takahashi}, \citenamefont
  {Katayama-Yoshida}, \citenamefont {Mochiku},\ and\ \citenamefont
  {Kadowaki}}]{terauchi1995electron}%
  \BibitemOpen
  \bibfield  {author} {\bibinfo {author} {\bibfnamefont {M.}~\bibnamefont
  {Terauchi}}, \bibinfo {author} {\bibfnamefont {M.}~\bibnamefont {Tanaka}},
  \bibinfo {author} {\bibfnamefont {T.}~\bibnamefont {Takahashi}}, \bibinfo
  {author} {\bibfnamefont {H.}~\bibnamefont {Katayama-Yoshida}}, \bibinfo
  {author} {\bibfnamefont {T.}~\bibnamefont {Mochiku}},\ and\ \bibinfo {author}
  {\bibfnamefont {K.}~\bibnamefont {Kadowaki}},\ }\bibfield  {title} {\bibinfo
  {title} {Electron-energy-loss spectroscopy of oxide superconductor bi2sr
  2cacu 2o 8},\ }\href@noop {} {\bibfield  {journal} {\bibinfo  {journal}
  {Japanese journal of applied physics}\ }\textbf {\bibinfo {volume} {34}},\
  \bibinfo {pages} {L1524} (\bibinfo {year} {1995})}\BibitemShut {NoStop}%
\bibitem [{\citenamefont {TERAUCHI}\ \emph {et~al.}(1999)\citenamefont
  {TERAUCHI}, \citenamefont {TANAKA}, \citenamefont {TSUNO},\ and\
  \citenamefont {ISHIDA}}]{terauchi1999development}%
  \BibitemOpen
  \bibfield  {author} {\bibinfo {author} {\bibnamefont {TERAUCHI}}, \bibinfo
  {author} {\bibnamefont {TANAKA}}, \bibinfo {author} {\bibnamefont {TSUNO}},\
  and\ \bibinfo {author} {\bibnamefont {ISHIDA}},\ }\bibfield  {title}
  {\bibinfo {title} {Development of a high energy resolution electron
  energy-loss spectroscopy microscope},\ }\href@noop {} {\bibfield  {journal}
  {\bibinfo  {journal} {Journal of microscopy}\ }\textbf {\bibinfo {volume}
  {194}},\ \bibinfo {pages} {203} (\bibinfo {year} {1999})}\BibitemShut
  {NoStop}%
\bibitem [{\citenamefont {Patel}\ and\ \citenamefont
  {Sachdev}(2019)}]{patel2019theory}%
  \BibitemOpen
  \bibfield  {author} {\bibinfo {author} {\bibfnamefont {A.~A.}\ \bibnamefont
  {Patel}}\ and\ \bibinfo {author} {\bibfnamefont {S.}~\bibnamefont
  {Sachdev}},\ }\bibfield  {title} {\bibinfo {title} {Theory of a planckian
  metal},\ }\href@noop {} {\bibfield  {journal} {\bibinfo  {journal} {Physical
  review letters}\ }\textbf {\bibinfo {volume} {123}},\ \bibinfo {pages}
  {066601} (\bibinfo {year} {2019})}\BibitemShut {NoStop}%
\bibitem [{\citenamefont {Thornton}\ \emph {et~al.}(2023)\citenamefont
  {Thornton}, \citenamefont {Liarte}, \citenamefont {Abbamonte}, \citenamefont
  {Sethna},\ and\ \citenamefont {Chowdhury}}]{thornton2023jamming}%
  \BibitemOpen
  \bibfield  {author} {\bibinfo {author} {\bibfnamefont {S.~J.}\ \bibnamefont
  {Thornton}}, \bibinfo {author} {\bibfnamefont {D.~B.}\ \bibnamefont
  {Liarte}}, \bibinfo {author} {\bibfnamefont {P.}~\bibnamefont {Abbamonte}},
  \bibinfo {author} {\bibfnamefont {J.~P.}\ \bibnamefont {Sethna}},\ and\
  \bibinfo {author} {\bibfnamefont {D.}~\bibnamefont {Chowdhury}},\ }\bibfield
  {title} {\bibinfo {title} {Jamming and unusual charge density fluctuations of
  strange metals},\ }\href@noop {} {\bibfield  {journal} {\bibinfo  {journal}
  {Nature communications}\ }\textbf {\bibinfo {volume} {14}},\ \bibinfo {pages}
  {3919} (\bibinfo {year} {2023})}\BibitemShut {NoStop}%
\end{thebibliography}%

\end{document}